\documentclass[twocolumn,apjl]{emulateapj}

\usepackage{color}

\def\mbf#1{\mbox{\boldmath ${#1}$}}

\begin{document}

\title{Protoplanetary Disk Winds by Magnetorotational Instability : \\
Formation of an Inner Hole and a Crucial Assist for Planet Formation}

\author{Takeru K. Suzuki$^{1,2}$, Takayuki Muto$^{3}$ 
\& Shu-ichiro Inutsuka$^{1,3}$}
\email{stakeru@nagoya-u.jp}
\altaffiltext{1}{Department of Physics, Nagoya University,
Nagoya, Aichi 464-8602, Japan}
\altaffiltext{2}{School of Arts and Sciences, University of Tokyo,
Komaba, Meguro, Tokyo, Japan, 153-8902}
\altaffiltext{3}{Department of Physics, Kyoto University, 
Kyoto, Japan, 606-8502}

\begin{abstract}
By constructing a global model based on 
3D local magnetohydrodynamical (MHD) simulations, 
we show that the disk wind driven by magnetorotational instability 
(MRI) plays a significant role in the dispersal of the gas component of 
proto-planetary disks. 
Because the mass loss time scale by the MRI-driven disk winds is proportional 
to the local Keplerian rotation period, a gas disk dynamically evaporates 
from the inner region with possibly creating a gradually expanding inner hole, 
while a sizable amount of the gas remains in the outer region.
The disk wind is highly time-dependent with quasi-periodicity of 
 several times Keplerian rotation period at each radius, which 
will be observed as time-variability of protostar-protoplanetary disk 
systems. 
These features persistently hold even if a dead zone exists because 
the disk winds are driven from the surface regions where ionizing cosmic 
rays and high energy photons can penetrate. 
Moreover, the predicted inside-out clearing significantly suppresses the 
infall of boulders to a central star and the Type I migration of proto-planets  
which are favorable for the formation and survival of planets.
\end{abstract}

\section{Introduction}
Planets are believed to form in protoplanetary disks  
consisting of gas and dust components around newly born stars. 
The evolution of the gas component is crucial in 
determining the final state of the planetary system, 
such as the number and locations of terrestrial (rocky) and 
Jovian (gas-giant) planets \citep{il04}. 
The amount of the gas that is captured by Jovian planets 
is generally much smaller than the total gas of the disk. 
Therefore, the gas component should dissipate via other mechanisms 
with the observationally inferred timescale of $10^6 - 10^7$ yr \citep[e.g.,][]
{hll01,her08}.
The currently favored scenario is that the gas dissipates by 
 the combination of photoevaporation by Ultraviolet (UV) flux 
 from a central star and viscous accretion \citep[e.g.][]
{shu93,mat03,tak05,alx06,cm07}.
However, the time-evolution of the luminosity and spectrum of 
the UV radiation is quite uncertain, 
and moreover some observed transitional disks with inner holes 
are inconsistent with the photoevaporation mechanism \citep{cal05,esp08,hug09}.

By performing local 3D ideal MHD simulations in 
the shearing box approximation \citep{hgb95,mt95}, 
\citet[][SI09 hereafter]{si09} have shown 
 that MHD turbulence in protoplanetary disks effectively drives 
 disk winds. 
The local timescale of 
the dynamical evaporation by magnetorotational instability (MRI) driven 
disk winds is defined as 
\begin{equation}
\tau_{\rm ev} =\Sigma/(\rho v_z)_{\rm w},  
\label{eq:evtm}
\end{equation}
where $(\rho v_z)_{\rm w}$ is the sum of the mass fluxes from the upper 
and lower disk surfaces and $\Sigma$ is the surface density.
We can estimate $\tau_{\rm ev} \sim$ several thousand years at 1 AU 
of the typical protosolar disk \citep[e.g., minimum mass solar nebula, or 
MMSN in short hereafter;][]{hay81};  here we use the typical 
values, $\Sigma = 1700$ g cm$^{-2}$ and $(\rho v_z)_{\rm w}\sim 10^{-8}$
g cm$^{-2}$s$^{-1}$ (see \S\ref{sec:ndzls} for the scaling of the disk wind 
flux). 
While the actual dispersal time is much longer as shown in this paper 
because the radial mass accretion is not taken into account here, 
this estimate shows that the MRI-driven disk wind should play a significant 
role in the dispersal of protoplanetary disks.  
In this paper, we investigate the evolution of gas density with disk winds 
and radial mass accretion in a global model. 

In SI09, we did not take into account the effects of a so-called dead zone 
which is inactive with respect to MRI because of the insufficient ionization 
for the coupling between gas and magnetic fields.   
The dead zone is believed to form around the midplane in the inner parts of
protoplanetary disks because ionizing cosmic rays and X-rays cannot penetrate 
from the disk surfaces (Sano et al.2000).    
In this paper, we improve the models of SI09 by investigating the effects of 
dead zones by performing resistive MHD simulations.

\begin{figure*}
\figurenum{1} 
\epsscale{0.9}
\begin{center}
\plotone{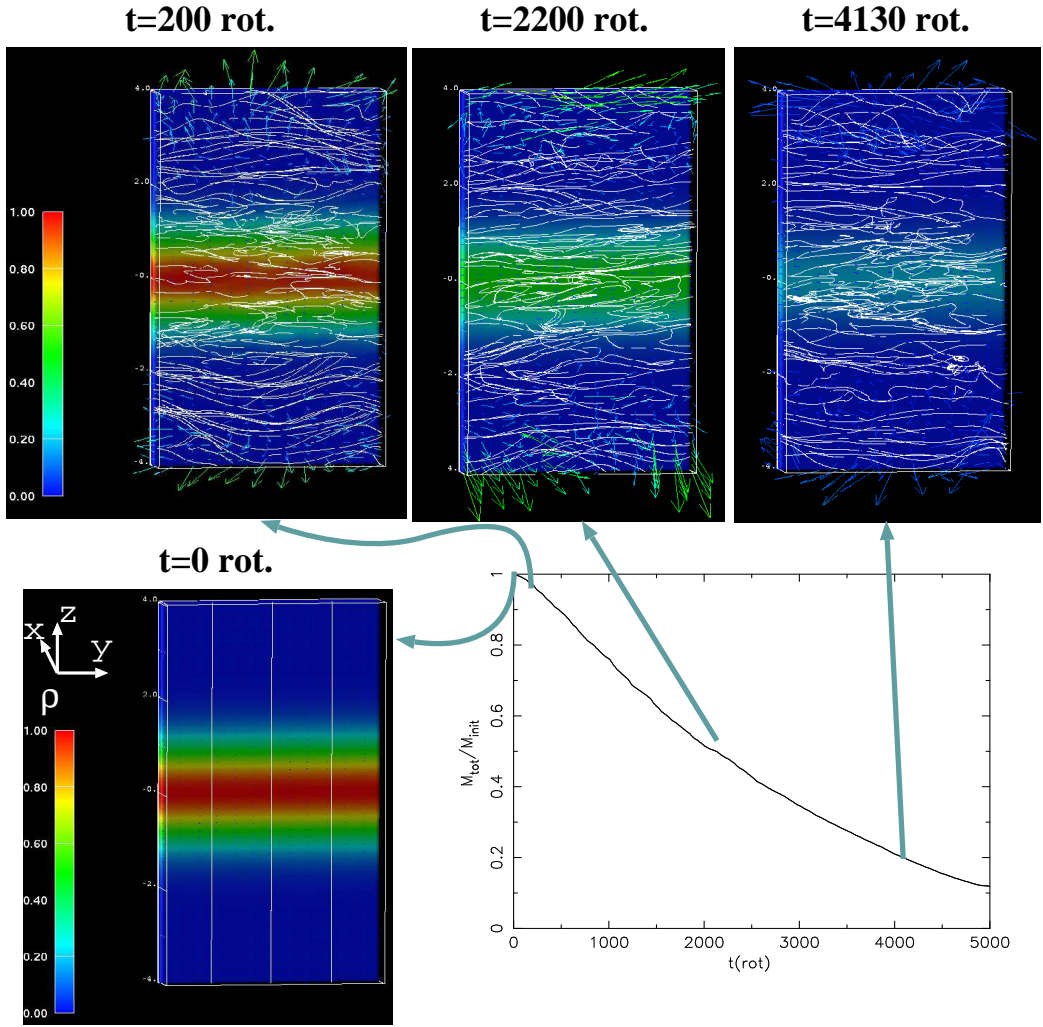}
\end{center}
\caption{
Dynamical evaporation of a protoplanetary gas disk by local 
3D ideal MHD simulation without mass supply by radial accretion. 
We impose weak vertical magnetic field, $B_z$, with the plasma $\beta$ 
value, $\beta_{z,{\rm mid}}=8\pi \rho_{\rm mid} c_s^2/B_z^2 = 10^6$, 
at the midplane.  
The lower right panel shows the total mass normalized by the initial mass 
in the local simulation box as a function of rotation time.
The four color panels show the snapshots of the local protoplanetary disk 
simulation at $t=0$ (initial condition), 200 rotations, 2200 
rotations, and 4130 rotations. The $x$, $y$, and $z$ components respectively 
correspond to radial, azimuthal, and vertical components. 
The unit of each component is scale height, $H\equiv \sqrt{2}c_s/\Omega$.  
The box size is $(x,y,z)=(\pm 0.5 H, \pm 2H, \pm 4H)$, which is resolved 
by (32,64,256) grid points. 
The colors indicate density 
normalized by the initial value at the midplane, the white solid lines 
illustrate magnetic field lines, and the arrows show velocity field.   
A small number of magnetic field lines (vertical lines) in 
the panel of $t=0$ reflect that the initially imposed magnetic field is weak 
(the number of field lines is scaled by magnetic field strength). 
We should emphasize that the actual dispersal is much slower because of the 
mass supply by accretion (see text).}
\label{fig:ttmsev}
\end{figure*}

The construction of the paper is as follows : 
We firstly describe the MHD simulations in local shearing boxes with 
and without dead zones in \S \ref{sec:lsb}.
In \S \ref{sec:mdl} we describe our global model. 
In \S \ref{sec:res}, we show how the MRI disk winds disperse the gas 
component of protoplanetary disks from the inner part. 
Also, we show its effects on the planet formation.     
In \S \ref{sec:dis}, we discuss the properties of disk evolution mainly 
focusing on the escape of the disk winds from the gravity of a central 
star. 

\section{Local Simulations}
\label{sec:lsb}
The main purpose of this section is to determine the turbulent viscosity 
and the mass flux of the disk winds which are used in the global models 
in \S \ref{sec:mdl}.  
For that purpose, we perform 3D MHD simulations of a local protoplanetary 
(accretion) disk in the shearing box coordinate \citep{hgb95}. 
We use the simulation box with $(x,y,z)=(\pm 0.5H, \pm 2H, \pm 4H)$, 
where the $x$, $y$, and $z$ components respectively correspond to radial, 
azimuthal, and vertical components and scale height, $H$, is defined 
from sound speed, $c_s$, and disk rotation frequency, $\Omega$, as 
$H\equiv \sqrt{2}c_s/\Omega$. 

In SI09 we have already shown results of ideal MHD simulation in the shearing 
box up to 400 rotation time. In this paper, we first extend this simulation 
until 5000 rotation time when the significant mass is lost from the 
simulation box (\S \ref{sec:ndzls}). 
In \S \ref{sec:dzls} we take into account the effects 
of resistivity ({\it i.e.} dead zone). Later in this paper, we perform 
simulations in boxes with larger vertical extents to study the effects of 
the box size on the escaping mass (\S \ref{sec:lslvb}).

\subsection{Launching of Disk Winds}
\label{sec:ons}
Before showing results of the local simulations, we discuss basic 
properties of the disk wind obtained in SI09 because the disk wind is a 
key that controls the evolution of protoplanetary disks in this paper. 
In SI09 we interpreted that the disk winds are driven by the breakups 
of channel-mode flows \citep[e.g.][]{san04} as a result of MRI \citep{bh91}. 
Large-scale channel flows \citep[e.g.][]{san04}
develop most effectively 
at 1.5 - 2 times the scale height above the midplane. 
The breakups of these channel flows by magnetic reconnections
drive disk winds in a time-dependent manner with quasi-periodic cycles of 
 5-10 rotation period. 
The disk material itself is lift up recurrently, which will be 
observed as the time variation of effective disk surfaces. 
Time-variabilities are actually observed in protostar-protoplanetary disk 
systems \citep[e.g.][]{wis08,muz09,bls09}, which might be explained by 
quasi-periodic breakups of channel flows.  
The quasi-periodic feature of the disk winds is universally found  
not only in ideal MHD simulations but in simulations with dead zones 
as will be shown in \S \ref{sec:dzls}.

We should note that upward motions in vertically stratified 
accretion disks have been widely discussed with various interpretations.
Magnetic buoyancy \citep[Parker instability;][]{par66} is 
one of the mechanism in uplifting mass and magnetic 
field in the upper regions with $|z|\gtrsim 1.5H$ \citep{ms00,mhm00,nmm06}. 
In fact, we observe $\frown$-shape magnetic field structures, which 
are characteristic of Parker instability, in our simulations as well 
(Figure \ref{fig:ttmsev}). Recently, \citet{lfg10} also discussed that 
channel flows are moving upward by magnetic buoyancy in a stratified disk. 
Magnetic buoyancy may play a role in upward flows and magnetic fields 
in cooperating with MRI channel flows in the upper regions ($|z|\gtrsim 1.5H$).
Detailed analysis of the role of magnetic buoyancy is driving the disk winds 
remains to be done. 

Flows and magnetic motions around the midplane show complicated behaviors.
On one hand, magnetic pattern seems to propagate away from the midplane 
\citep[][; our simulation shows the same trend.]{dav10,gre10}. 
Since this region is buoyantly stable \citep{skh10}, this apparent 
propagating pattern may be due to a different mechanism from magnetic buoyancy. 
\citet[][see also Gressel 2010]{bra95} try to interpret 
this upgoing pattern from dynamo waves \citep{par55,yos75,vb97}.
On the other hand, the direction of the Poynting flux associated with 
magnetic tension is toward the midplane as we have shown in SI09. 
We explained that 
the breakups of large-scale channel flows at injection regions, 
$z\approx \pm 1.5H$, drive mass motions to a midplane as well as
surfaces. 

Although properties of vertical flows and magnetic motions are not 
clearly understood in detail, upward flows in $z\gtrsim$ 
a few scale heights seem to be robust phenomena in stratified disks.  
The mass loading at the injection regions appear to be operated by the 
breakups of channel flows of MRI, and magnetic buoyancy also plays a role 
in lifting up the gas in the upper regions.


\subsection{Disks without Dead Zone}
\label{sec:ndzls}

We show results of the ideal MHD simulations in this subsection. 
We adopt the same resolution of the grid points, $(x,y,z)=(32,64,256)$, 
as in SI09 for runs with different net vertical magnetic flux. 
In addition to the standard resolution runs, we perform the simulations 
with higher resolution, $(x,y,z)=(64,128,512)$, in some cases. 
Extending from SI09, we carry out a long-time simulation in this paper 
until the significant mass is lost from the local box by the disk wind. 
Figure \ref{fig:ttmsev} is the result of the initial plasma 
$\beta$ value, $\beta_{z,{\rm mid}} = 8\pi \rho_{\rm mid} c_{\rm s}^2/B_z^2 
=10^6$, at the midplane for net vertical magnetic 
field\footnote{Note that the integrated 
vertical magnetic flux, $\int dx dy B_z$ is a strictly conserved quantity in 
the local shearing box simulations \citep{hgb95}, though the magnetic energy 
of $z$ component $\int dx dy B_z^2/4\pi$ does not conserve. 
Then, $\beta_{z,{\rm mid}}$ 
is a good indicator of magnetic field strength not only at the initial state 
but at later times when the magnetic field is amplified by MRI.}, $B_z$, 
where $\rho_{\rm mid}$ is the density at the midplane. 
The figure shows that about 90\% of the initial 
gas is dispersed in 5000 rotations (5000 years at 1 AU) if there is no mass 
supply by accretion.
This result shows that the evaporation timescale using Equation 
(\ref{eq:evtm}) gives a reasonable estimate and the MRI-driven disk winds 
should significantly affect the evolution of protoplanetary disks. 

\begin{figure}
\figurenum{2} 
\epsscale{1.}
\begin{center}
\plotone{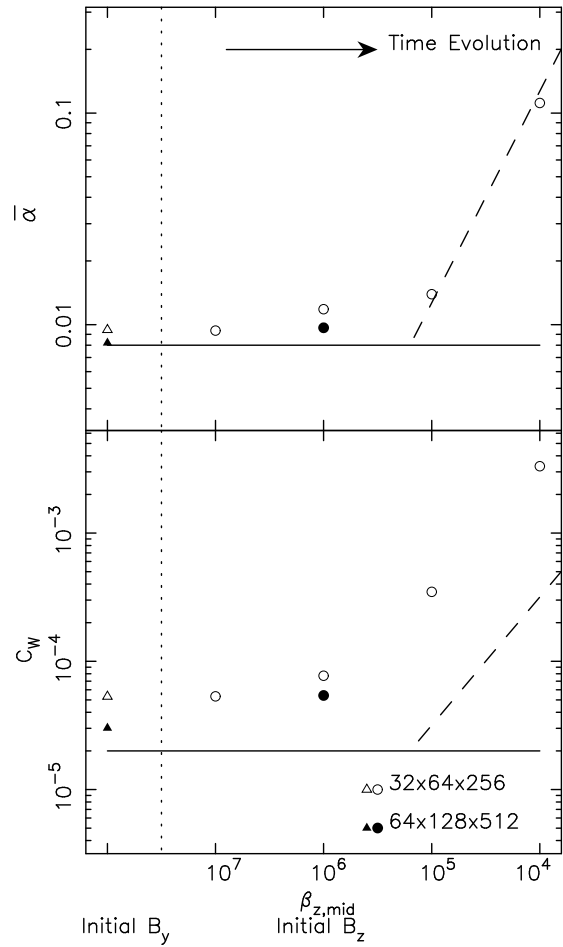}
\end{center}
\caption{Dependences of averaged turbulent viscosity, $\bar{\alpha}$, (top) 
and mass flux of disk winds, $C_{\rm w}$, (bottom)
on 
the initial plasma $\beta_{z,{\rm mid}}(=8\pi\rho_{\rm mid} c_s^2/B_z^2)$ 
values for the net vertical field, $B_z$. 
The left-most grid corresponds to the initial toroidal field cases. 
The open symbols are the results of the standard resolution simulations 
with mesh points $(x,y,z)=(32,64,256)$ and the filled symbols are the results 
of the higher resolution of $(64,128,512)$.
The solid lines represent $\bar{\alpha_{\rm fl}}$ and $C_{\rm w,fl}$, which 
we use for the global model. The dashed lines denote the linear dependence 
on $\beta_{z,{\rm mid}}^{-1}$.} 
\label{fig:btdep}
\end{figure}

From the results of the local MHD simulations, we can determine the Shakura 
\& Sunyaev (1973) $\alpha$ viscosity and the mass flux of the disk winds. 
$\alpha$ is calculated from anisotropic stress of MHD turbulence, 
\begin{equation}
\alpha=(v_x\delta v_y - \frac{B_x B_y}{4\pi\rho})/c_s^2, 
\label{eq:alp}
\end{equation}
where $\delta v_y \equiv v_y + (3/2)\Omega x $ is the velocity shift from 
the background Kepler rotation ($3/2)\Omega x$). 
We use the average $\alpha$ in the entire simulation box, 
\begin{equation}
\bar{\alpha} = \frac{\int \rho \alpha dx dy dz}{\int \rho dx dy dz}. 
\end{equation} 
Note that the density weighted average, $\bar{\alpha}$, is directly related 
to mass accretion rate \citep[e.g.,][]{pri81}.  
As for the disk wind mass flux, we 
use the nondimensionalized mass flux, 
\begin{equation}
C_{\rm w}=(\rho v_z)_{\rm w}/(\rho_{\rm mid}c_s).
\label{eq:ndms}
\end{equation}
Figure \ref{fig:btdep} shows $\bar{\alpha}$ and  $C_{\rm w}$ as functions of 
$\beta_{z,{\rm mid}}$. 
Note that the net $B_z$ becomes strong from left to right. The left-most 
grid is for the cases of no net $B_z$ field. In these cases we initially 
give purely toroidal field, $B_y$, with the $\beta$ values of $10^6$ in 
$-3H < z < 3H$, whereas the results do not depend on the initial strength 
because magnetic flux of $y$ (as well as $x$) component does not conserve.
  
The figure exhibits that both $\bar{\alpha}$ and $C_{\rm w}$ have the 
floor values for sufficiently weak net vertical magnetic field, 
$\beta_{z,{\rm mid}}\gtrsim 10^6$. 
In these cases the saturated magnetic field at the midplane 
gives $\beta (=8\pi p/B^2) \sim 100$, indicating $\approx 1$\% of the gas 
energy is transferred to the magnetic fields (SI09). The saturation level 
roughly corresponds to the floor value of $\alpha \sim 0.01$ ($\alpha$ is 
roughly $1/\beta$). 
Such weak net vertical field gives little effects, because it is much smaller 
than the turbulent component of the vertical fields. 
The turbulent component of $B_z$ gives $\langle B_z^2\rangle/8\pi 
\sim 10^{-4}-10^{-3}p$ even in the zero 
net vertical field case, where $\langle \rangle$ denotes time-average. 
This value seems to determined as the level that is one or two orders of 
magnitude smaller than the dominant toroidal component 
($\langle B_y^2\rangle/8\pi\sim p/\beta \sim 0.01 p$).
If the net vertical field is much smaller than the turbulent vertical 
component, $\langle B_z \rangle^2/8\pi \le 10^{-6}p$
the values of $\alpha$ and $C_{\rm w}$ are not affected by the net vertical 
field. 

On the other hand, for larger magnetic field cases, $\beta_{z,{\rm mid}}
\lesssim 10^5$, the net vertical magnetic field plays a role, because 
it is not negligible compared to the turbulent component. $\bar{\alpha}$ and 
$C_{\rm w}$ increase almost linearly with magnetic energy of net vertical 
field ($\propto 1/\beta_{z,{\rm mid}}$). 
The behaviors of $\bar{\alpha}$ and $C_{\rm w}$ are similar, which indicates 
that the mass flux of the disk winds is positively correlated with the mass 
accretion rate. 
This is reasonable because the energy source of the disk winds 
is the gravitational energy liberated by accretion, which is discussed in 
\S \ref{sec:dis}. 

In the higher resolution runs, the saturation levels of the magnetic fields 
are slightly lower than the corresponding cases with lower resolutions. 
Then, both $\bar{\alpha}$ and $C_{\rm w}$ become slightly smaller in the higher 
resolution runs. The dependence of the saturation level of magnetic 
field and $\alpha$ on resolutions is still under debate and widely discussed 
in various authors \citep[e.g.][]{bh98,pcp07}

\subsection{Disks with Dead Zone}
\label{sec:dzls}
The temperature of protoplanetary disks are too low to ionize 
the gas by thermal collisions. Various ionization sources, such as X-rays, 
cosmic rays, and radioactive nuclei, have been widely discussed 
\citep{un80,hay81,gni97}. 
The ionization degree at midplanes 
is generally smaller than that in upper regions because the recombination 
rate is higher there as a result of higher density.
 Under certain 
circumstances, dead zones, in which the gas is decoupled with the magnetic 
fields due to the insufficient ionization, are supposed to form near 
midplanes. 
However, \citet{is05} also introduced a self-sustained mechanism by 
current-carrying electrons in turbulent disks, which increases the resultant 
ionization degree. 
Thus, calculating ionization degree is not straightforward.    
In this paper, we study extreme cases which give distinct dead zones. 
As will be described below, we take into account the ionization by cosmic 
rays and X-rays. Since they come from disk surfaces, dead zones tend to 
form near the midplane if the surface density is sufficiently high.  
We adopt a simple treatment in determining ionization degree 
and study effects of dead zones on disk winds. 

\subsubsection{Set-up}

To treat dead zones, we take into account resistivity in the induction 
equation that describes the evolution of magnetic fields :   
\begin{equation}
\frac{\partial \mbf{B}}{\partial t} = \mbf{\nabla} \times 
(\mbf{v}\times \mbf{B} - \eta \mbf{\nabla}\times \mbf{B}) ,
\end{equation}
where $\eta$ is resistivity which is determined by ionization degree, 
$x_{\rm e}$ as $\eta = 234 \sqrt{T} /x_e$ \citep{bb94},  
because electrons control the coupling between gas and magnetic field 
in high density medium.
We assume the temperature structure of the MMSN 
\citep{hay81} : 
\begin{equation}
T = 293{\rm K}\left(\frac{r}{1\;{\rm AU}}\right)^{-1/2} , 
\label{eq:tmprt}
\end{equation}
which can be transformed into the sound speed, 
$c_s = 0.99{\rm km/s}\left(\frac{r}{\rm 1 \; AU}\right)^{-1/4}$.
\begin{figure}[h]
\figurenum{3} 
\epsscale{0.8}
\begin{center}
\plotone{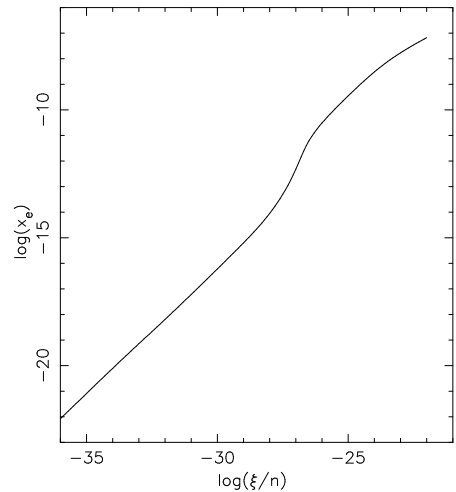}
\end{center}
\caption{Ionization degree, $x_e$, on ionization rate per number density, 
$\xi/n$. The data are taken from \citet{is05,Sano00}. 
}
\label{fig:iond}
\end{figure}

The ionization degree is determined by the balance between ionization and 
recombination. 
In this paper, we use 
the result of previous calculation by \citet{Sano00} and \citet{is05}, 
which give $x_e$ as a function of ionization rate, $\xi$, and local number 
density, $n$, by calculating recombination on dust grains and radiative 
and dissociative recombination in gas phase \citep[see also, e.g.][]
{un80,in06}. 
When one fixes the abundance of gas and the properties of dust grains, 
the ionization degree is a function of the only one parameter, $\xi/n$ 
\citep{is05}. 
This is because the recombination is essentially two-body reaction and 
recombination rate per volume is proportional to $n^2$ while the 
ionization rate per volume is $\xi n$.  
Figure \ref{fig:iond} presents $x_e$ for the solar abundance gas with 
gas-to-dust ratio of 100 and dust grain size of $0.1 \mu$m \citep{is05,Sano00}.
We use this data for our local resistive MHD simulations. 
We adopt the mean molecular weight, $\mu = 2.3$, 
for the conversion between $n$ and $\rho$. 
This $\mu$ value reflects the condition that the major component is H$_2$ 
molecules.

We take into account the ionization by Galactic cosmic rays and X-rays from 
a central star. 
The total ionization rate, $\xi$, is the sum of these two sources, 
\begin{equation}
\xi = \xi_{\rm CR}+\xi_{\rm X},
\end{equation}
where subscripts ${\rm CR}$ and ${\rm X}$  
represent cosmic rays and stellar X-rays, respectively.
We adopt $\xi_{\rm CR} = 10^{-17}\exp(-l/l_{\rm cr})$ s$^{-1}$, 
where $l$ is the column densities 
integrated from the upper or lower surfaces and $l_{\rm cr}=100$ g cm$^{-2}$ 
is the path length of cosmic rays \citep{hay81,tsd07}.

\begin{figure}[h]
\figurenum{4} 
\epsscale{1.1}
\begin{center}
\plotone{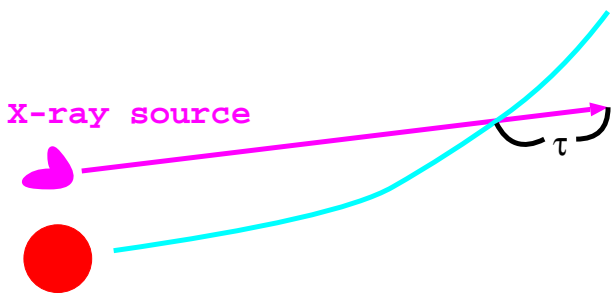}
\end{center}
\caption{The path, $ds$, of emitted X-ray from the source located 
above a central star to the disk (pink line). The blue line is the 
upper disk surface. (see text)}
\label{fig:dzct}
\end{figure}

Observations of T-Tauri stars show high X-ray activities. The typical 
luminosity in X-ray wavelength is $10^{29-31}$ erg s$^{-1}$, which is 
3-5 orders of magnitude higher than the level of the present Sun 
\citep[e.g.][]{tel07}. 
These X-rays are supposed to be an efficient ionization source.  
Following \citet{gni97}, 
we model the ionization rate, $\xi_{\rm X}$, by the X-rays.
We assume the X-ray sources located at 10 $R_{\odot}$ ($\sim 3-5$ 
stellar radii) above and below a star, where $R_{\odot}$ denotes the 
solar radius. The X-ray follows the path shown in Figure \ref{fig:dzct}.  
The ionization rate is obtained by \citet{gni97,ftb02}
as follows: 
$$
\hspace{-3cm}\xi_{\rm X} / J(\tau) = 1.2\times 10^{-11} ({\rm cm^{-2}s^{-1}})
$$
\begin{equation}
\times \left(\frac{L_{\rm X}}{10^{30}
{\rm erg \; s^{-1}}}\right)\left(\frac{r}{1{\rm AU}}\right)^{-2}\left(
\frac{\sigma}{4\times 10^{-24}{\rm cm^{2}}}\right) , 
\end{equation} 
where $L_{\rm X}$ is the X-ray luminosity, $\sigma \approx 4\times 10^{-24}
\left(\frac{E_{\rm X}}{3{\rm keV}}\right)^{-2.81}$ is 
the absorption cross section depending on the X-ray energy, $E_{\rm X}$, 
$\tau=\int ds n \sigma$ is optical depth and 
$J(\tau) = 0.686 \tau^{-0.606}\exp(-1.778\tau^{0.262})$. 
Here, the optical depth, $\tau$, is integrated along the path, $ds$, in Figure 
\ref{fig:dzct} for the variation of scale height, $H\approx 0.05{\rm AU}\left(
\frac{r}{1\;{\rm AU}}\right)^{5/4}$. 

\begin{figure}
\figurenum{5} 
\epsscale{1.}
\begin{center}
\plotone{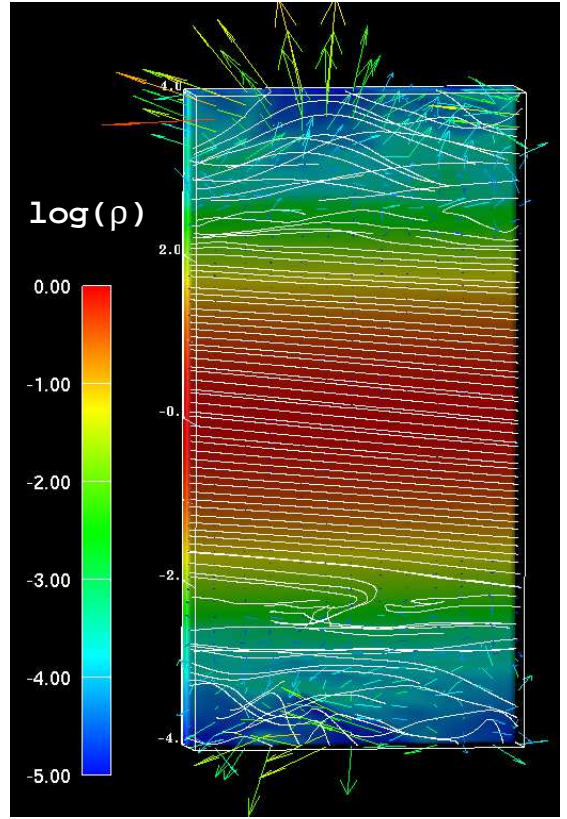}
\end{center}
\caption{Snap-shot of the local disk with dead zone at 
250 rotations in the case with medium X-ray activity at 1 AU.
The colors indicate the {\it logarithmic} scale of density, the solid lines 
represent magnetic fields, and the arrows are velocity field.}
\label{fig:snpdz1}
\end{figure}

\begin{figure*}
\figurenum{6} 
\epsscale{0.9}
\begin{center}
\plottwo{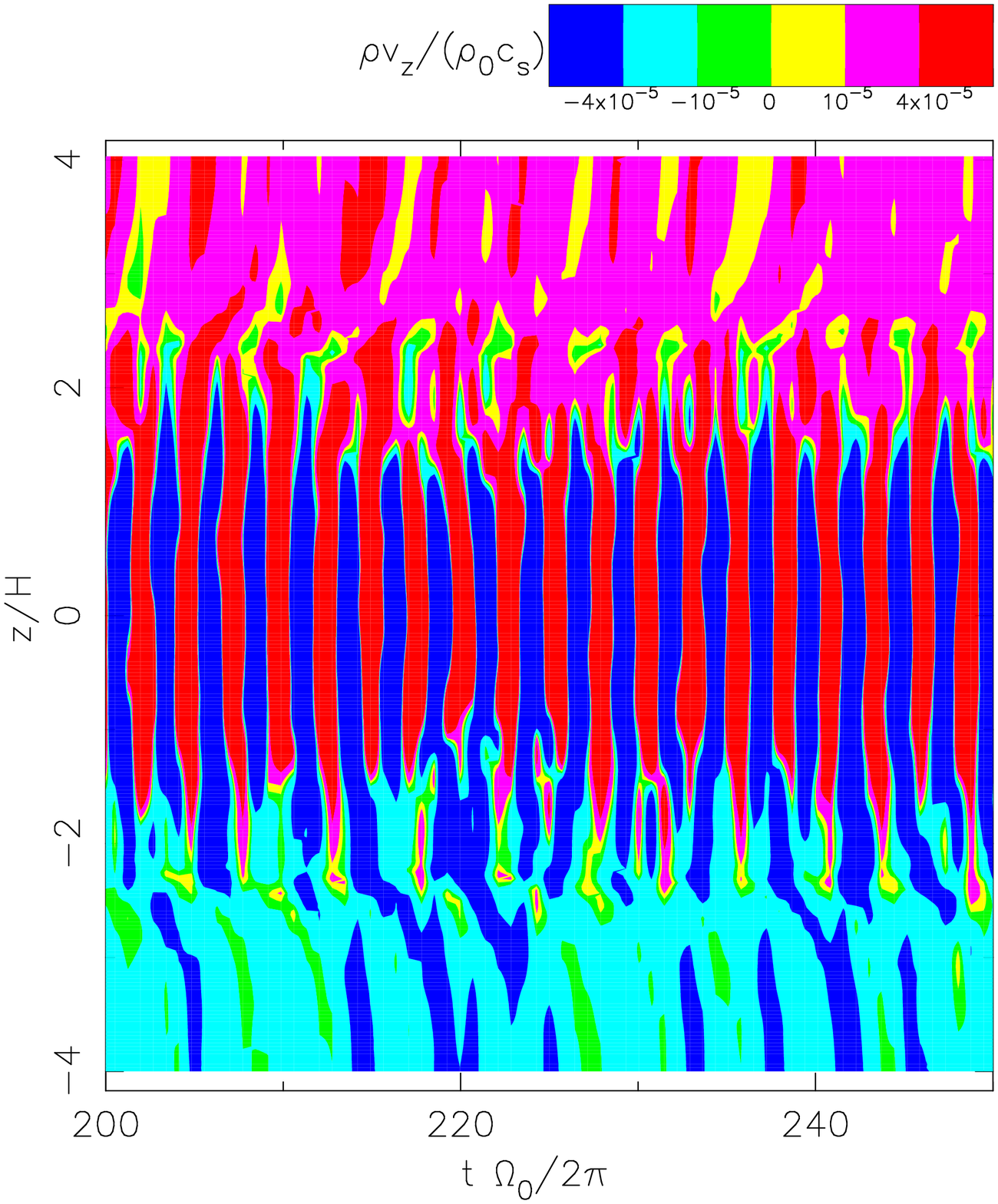}{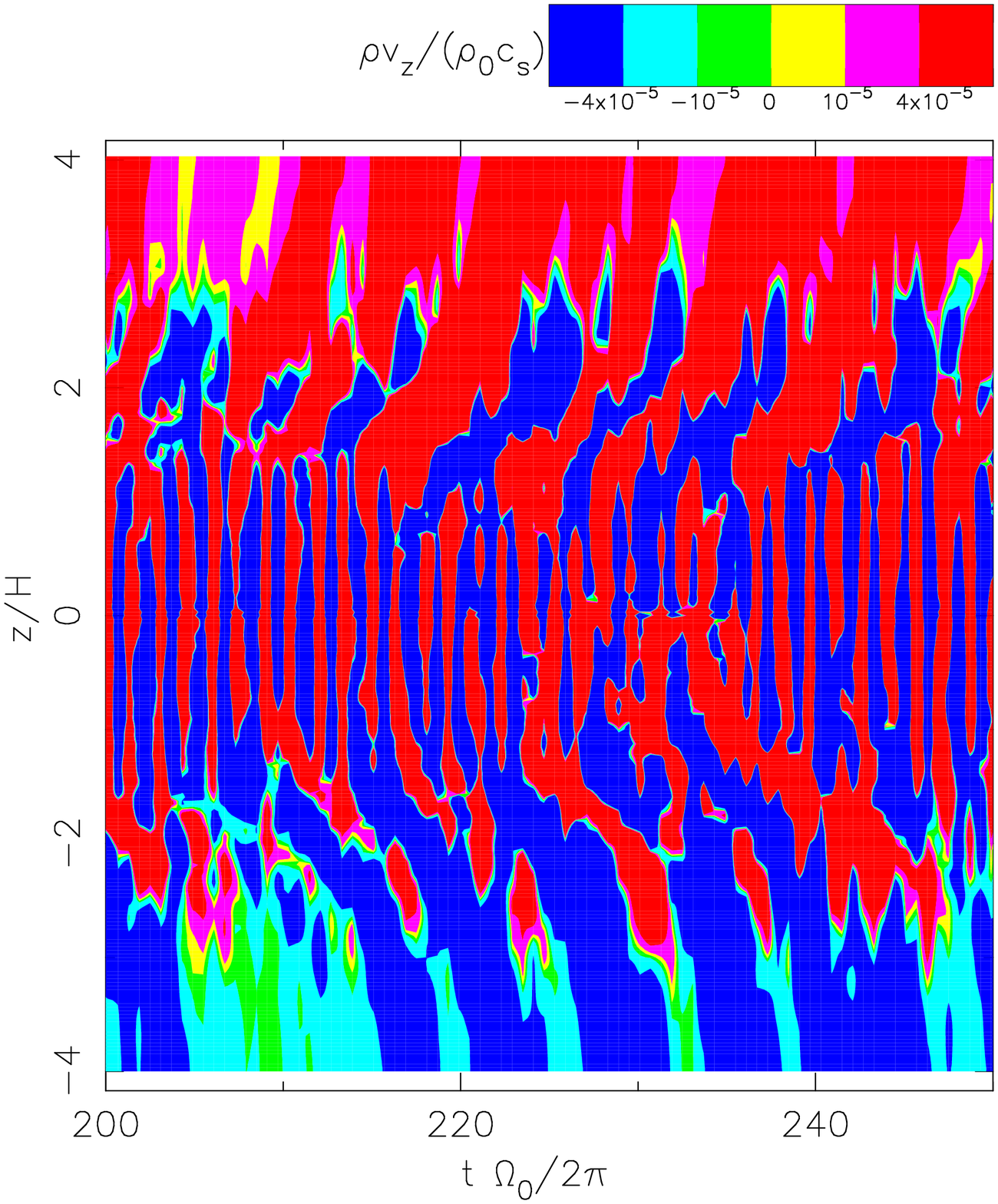}
\end{center}
\caption{Comparison of time-height diagram of $\rho v_z$ between the resistive 
MHD case of the medium X-ray ($L_{\rm X} = 10^{30}$ erg s$^{-1}$ and 
$E_{\rm X}=3$ keV) at 1 AU (left panel) and the ideal MHD case (right 
panel). $\rho v_z$ is averaged over the $x-y$ planes.
The horizontal axis is in unit of rotation period, and the vertical axis 
is normalized by scale height, $H=\sqrt{2}c_s/\Omega$. }
\label{fig:rvz1}
\end{figure*}

\begin{figure}
\figurenum{7} 
\epsscale{1.2}
\begin{center}
\plotone{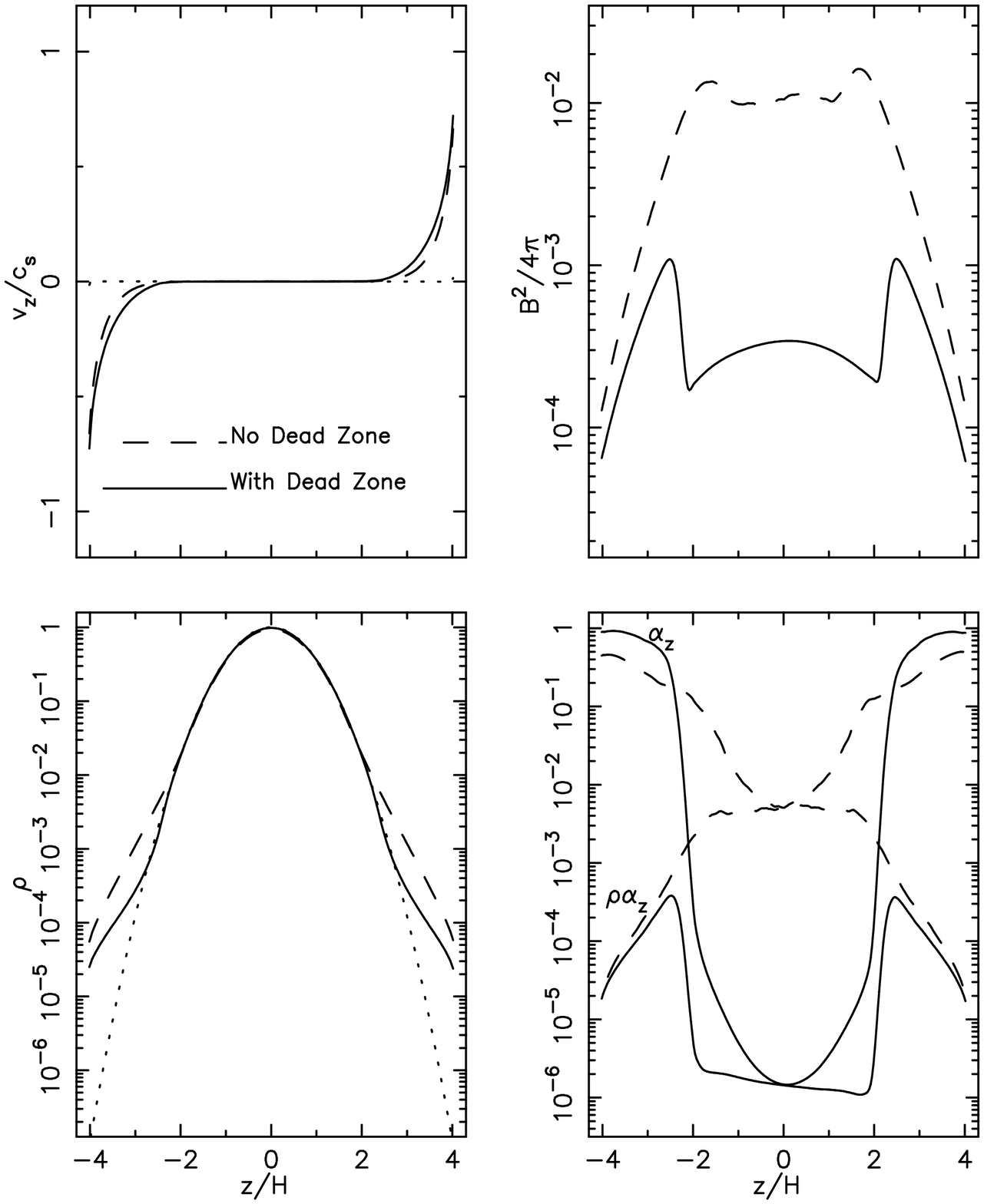}
\end{center}
\caption{Comparison of the resistive MHD (dead zone) case of the medium X-rays 
at 1AU (solid lines) 
with the ideal MHD (no dead zone) case (dashed lines). 
Both cases assume the net vertical magnetic field with $\beta_{z,{\rm mid}} 
= 10^6$. The horizontal axis is vertical height in unit of 
$H(=\sqrt{2}c_s/\Omega)$. The upper left panel shows the vertical velocities 
normalized by the sound speed; the lower left panel shows the densities; 
the upper right panel shows the magnetic energy. The lower right panel 
shows $\alpha_z$ and $\rho \alpha_z$. 
}
\label{fig:dz1}
\end{figure}

We perform simulations at three different locations, $r=1, 5, 25$ AU. 
We consider the three levels of X-ray activities : 
$(L_{\rm X}({\rm erg\;s^{-1}}),E_{\rm X}({\rm keV})) = (10^{31},5)$, 
$(10^{30},3)$, and $(10^{29},1)$, which we call the strong, medium, and weak 
X-ray cases, respectively. 
We adopt the original MMSN for the surface density, $\Sigma = 1700 
\;{\rm g\;cm^{-2}}\left(\frac{r}{1\; {\rm AU}}\right)^{-3/2}$. 
In the resistive MHD simulations we adopt the only standard resolution  
($(x,y,z)=(\pm 0.5 H, \pm 2H, \pm 4H)$ is resolved by $(32,64,256)$ grid 
points).
We impose weak net vertical magnetic field, $\beta_{z,{\rm mid}} =10^6$, 
at the midplane. 
The constant $\beta_{z,{\rm mid}}$ distribution indicates that we assume the 
dependence of the net vertical field of $\langle B_z\rangle\approx 0.01\;{\rm G}
\left(\frac{r}{1{\rm AU}}\right)^{-13/8}$ for the MMSN
($\rho_{\rm mid} \propto r^{-11/4}$ and $c_s\propto r^{-1/4}$).

\subsubsection{Result}
First, we describe results of the case with the medium X-ray activity 
at 1 AU in detail before discussing results of the different locations 
and X-ray activities. 
Figure \ref{fig:snpdz1} shows a snapshot of the local disk structure 
at 250 rotations.  
(Note that the density is in {\it logarithmic} scale here; {\it c.f.}, 
a linear scale was used in Figure \ref{fig:ttmsev}.)
The figure shows that the magnetic field lines become almost straight in 
the dead zone, $-2H \lesssim z \lesssim 2H$, because MRI does not operate 
due to the insufficient ionization here. This is a clear contrast to 
disks without dead zones (Figure \ref{fig:ttmsev}; see also SI09). 
In the surface regions, however, the magnetic fields are more turbulent 
because the ionizing cosmic rays and X-rays can penetrate to these regions 
and MRI is active at $r\approx \pm(2-3)H$. 
One can also see that the disk winds stream out of both surfaces because the 
disk winds are driven from the surface regions with sufficient ionization 
rather than a deeper midplane location.
The breakups of the channel flows triggered by MRI at $z\approx 2H$ and 
Parker instability in $z\gtrsim 3H$ drive these disk winds, as explained 
in \S\ref{sec:ons}. For example, one can see a typical 
$\supset$-shape channel flow structure at $z\simeq -2H$, and a 
$\frown$-shape structure at $z\simeq (3-4)H$, which is typical for Parker 
instability.   
Although the mass flux of the disk winds become moderately smaller than the
ideal MHD run, the dead zone give little effects on the disk winds 
(see below).  

Figure \ref{fig:rvz1} compares the time-height diagrams of the mass flux, 
$\rho v_z$ of the same case 
(left panel)
to the result of the ideal MHD case (right panel). 
$\rho v_z$ is averaged over the $x-y$ planes at each time step. 
The left panel shows that disk winds flow out of the upper and 
lower surfaces even though the dead zone forms around the midplane.  
This is because the disk winds are excited from the surface regions, 
which we called injection regions in SI09, 
with the sufficient ionization, rather than deeper locations near the midplane. 
These injection regions are a consequence of the breakups of large-scale 
channel flows. 
The disk winds flow out intermittently with quasi-periodicity of 5-10 
rotations owing to the quasi-periodic breakups of the channel flows. 
The recurrent nature of the disk winds universally holds in both cases 
with and without dead zones, 
which might explain observed time-variabilities of 
protostar--protoplanetary disk systems \citep{wis08,muz09,bls09}. 
The heights of the injection regions become slightly higher than those 
of the ideal MHD case because the ionization degree at the deeper 
locations is not sufficient for MRI. Accordingly the mass flux of disk 
winds from the dead zone case is a little smaller than that of the no dead 
zone case.       

Figure \ref{fig:dz1} compares the vertical structure of these cases averaging 
over 200 rotations after the quasi-steady states are achieved. 
$\alpha_z$ in the lower right panel is the average of 
$\alpha$ (Equation \ref{eq:alp}) on each $x-y$ plane. 
The right panels (magnetic energy and $\alpha$ values) show that the dead 
zone extends from $z=-2H$ to $2H$ in the resistive MHD case. 
$\alpha_z$ sharply declines at $z=\pm 2H$ from the surface regions toward 
the midplane. It is expected that the mass accretion mainly takes place 
in the surface regions \citep{gam96}. 
The averaged $\alpha$ value, $\bar{\alpha}(=\int dz \rho \alpha_z/\Sigma)$, 
which directly determines global mass accretion rate, of the dead zone case 
is $\sim 3 \times 10^{-4}$, while it is $\approx 0.011$ in the no dead zone 
case (Figure \ref{fig:btdep}).  
On the other hand, in the surface regions the magnetic energy and $\alpha_z$ 
of the resistive MHD case are similar to those of the ideal MHD case because 
the sufficient ionization is achieved there by the X-rays and cosmic rays 
from the surfaces. 
The disk winds are effectively accelerated from the injection regions, and 
the velocity structures are very similar in both cases (the upper left panel 
of Figure \ref{fig:dz1}). The density of the winds are smaller because 
the heights of the injection regions are slightly higher in the dead zone 
case. Accordingly, the mass flux of disk winds of the resistive MHD 
case is the half of the mass flux of the ideal MHD case.

\begin{figure}
\figurenum{8} 
\epsscale{1.}
\begin{center}
\plotone{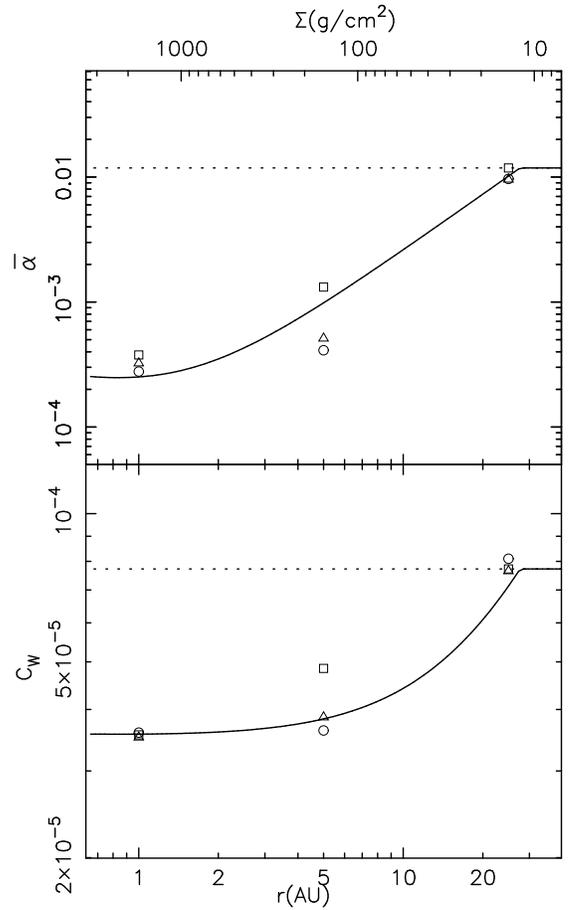}
\end{center}
\caption{Turbulent viscosity, $\bar{\alpha}$,  (top) and disk wind mass flux, 
$C_{\rm} (=(\rho v_z)_{\rm w}/(\rho_{\rm mid}c_s))$, (bottom) 
of the resistive MHD simulations at different locations. The squares 
are the results of the strong X-ray cases ($L_{X}=10^{31}$ erg s$^{-1}$, 
$E_x=5$ keV), the triangles are the results of the medium X-ray cases 
($L_X=10^{30}$ erg s$^{-1}$ and $E_X = 3$ keV), and the circles are the 
results of the weak X-ray cases 
($L_X=10^{29}$ erg s$^{-1}$ and $E_X = 1$ keV). The solid lines are the
fitting formula (Equations \ref{eq:dzal} and \ref{eq:dzcw}) used for the 
global calculations. The dotted lines are the results of ideal MHD 
simulations with net vertical field, $\beta_{z,{\rm mid}}=10^6$. }
\label{fig:dz2}
\end{figure}


Figure \ref{fig:dz2} summarizes the averaged turbulent viscosities, 
$\bar{\alpha}$, and the mass fluxes of the disk winds for the different 
locations and X-ray activities. 
The figure illustrates that both $\bar{\alpha}$ and $C_{\rm w}$
are smaller in the inner regions.
Since surface density is larger for smaller $r$, ionizing cosmic rays 
and X-rays cannot reach the midplane in the inner disk. A large 
fraction is occupied by the dead zone in the inner parts of disks. 
The $\bar{\alpha}$ values are reduced by more than an order of magnitude at 
$r=1$ AU. 
On the other hand, $C_{\rm w}$ is not so reduced, being half at most at 
$r=1$ AU, because the disk winds are excited from the surface regions 
and not so affected by the dead zones.  

\vspace{1cm}

\section{Global Modeling}
\label{sec:mdl}


We solve the time-evolution of surface density with mass accretion and 
disk wind mass loss \citep[e.g.][]{alx06}: 
\begin{equation}
\frac{\partial \Sigma}{\partial t} - \frac{1}{r}\frac{\partial}{\partial r}
\left[\frac{2}{r\Omega}\frac{\partial}{\partial r}(\Sigma r^2 \alpha c_s^2)
\right] + (\rho v_z)_{\rm w} = 0 .
\label{eq:sgmev}
\end{equation}
The second term denotes the radial mass flow by turbulent viscosity.
Here and from now, we simply write $\alpha$ for the turbulent viscosity 
in the global models, which is adopted from $\bar{\alpha}$ of the shearing 
box simulations of the previous section. 
The third term is the mass loss by disk winds, where
we here assume the specific angular momentum carried 
in the disk winds is the same as that in the disk material. 
Both $\alpha$ and $(\rho v_z)_{\rm w}$ 
are adopted from the local 3D MHD simulations in \S \ref{sec:ndzls} and 
\ref{sec:dzls}.

We assume the temperature structure of the MMSN (Equation \ref{eq:tmprt}). 
The initial surface density profile is also adopted from the MMSN, 
\begin{equation}
\Sigma = f_{\rm g}\Sigma_0\left(\frac{r}{\rm 1\; AU}\right)^{-3/2} 
\exp(-r/r_{\rm cut}), 
\label{eq:sgmmsn}
\end{equation}
where $\Sigma_0=2400$ g cm$^{-2}$ at 1 AU ($f_{\rm g}=0.7$ for the original 
Hayashi MMSN) and we use a cut-off radius, $r_{\rm cut} = 50$ AU. 
Although in the local resistive MHD simulations for dead zones 
(\S \ref{sec:dzls}) we adopted the original MMSN value, $f_g \Sigma_0 
=1700\;{\rm g\; cm^{-2}}$, 
a specific choice of $f_g$ in the local simulations does not change
the results of the global disk because we explicitly take into account 
the dependences of $\alpha$ and $C_{\rm w}$ on $\Sigma$ (\S \ref{sec:dzmd}).  

We integrate Equation (\ref{eq:sgmev}) by using the nondimensionalized 
variables in unit of $r_0 = \Omega_0 = f_{\rm g}\Sigma_0  = 1$, so the scaling 
factor, $f_{\rm g}$, 
does not appear explicitly in the calculations.  
The calculation region\footnote{Our results 
are not affected by the location of $r_{\rm in}$. The adopted value, 
$r_{\rm in}\approx 2R_{\odot}$, roughly coincides a typical radius 
of T-tauri stars, where $R_{\odot}$ is the solar radius. 
In realistic situations, however,  
a disk may truncate at several stellar radii where the Keplerian rotation 
frequency equals to the rotation frequency of the corotating stellar 
magnetic field \citep{gl79}. 
The matter can directly accrete to a central star through connecting 
flux tubes \citep{ken96}. } 
is from $r_{\rm in}=0.01$ AU
to $r_{\rm out} = 10000$ AU which is resolved by 2000 mesh points with 
grid spacing in proportion to $\sqrt{r}$.

\subsection{Disks without Dead Zone}
We apply the $\alpha$ and $C_{\rm w}$ obtained in the local ideal MHD 
simulations (\S \ref{sec:ndzls} and Figure \ref{fig:btdep}) to the global 
model. 
Figure \ref{fig:btdep} can be regarded as a time sequence along with disk
evolution because the surface density of protoplanetary disks decreases 
while vertical magnetic flux is supposed to be rather kept 
constant. 
If initial vertical magnetic fields are zero or very weak 
($\beta_{z,{\rm mid}}\gg 10^7$), $\alpha$ and $C_{\rm w}$ are expected to stay 
almost constant even after surface density decreases considerably. 
For such situations, we use constant  
$\alpha = \alpha_{\rm fl} = 8\times 10^{-3}$ and $C_{\rm w} = C_{\rm w,fl} = 
2\times 10^{-5}$ (solid lines in Figure \ref{fig:btdep}), 
which are the floor values obtained by the local simulations. 
For the disk wind mass flux, we choose a conservative value because 
the actual mass flux might be moderately smaller by returning mass from higher 
altitudes (\S \ref{sec:eng} and \ref{sec:esc}). 

When $C_{\rm w}$ stays constant, the disk wind flux has the following scaling :
\begin{equation}
(\rho v_z)_{\rm w} = C_{\rm w} \rho_{\rm mid}c_s\propto \Sigma r^{-3/2} ,
\label{eq:msfx}
\end{equation} 
where for the last proportionality we use $\rho_{\rm mid}c_s \propto 
\Sigma \Omega$ and assume a Keplerian rotating disk . 
Equation (\ref{eq:msfx}) shows that the wind mass flux is larger for smaller 
$r$, and the dispersal of protoplanetary disks by the disk winds starts 
from the inner part. 

If initial vertical fields are not so weak, $\alpha$  
and $C_{\rm w}$ eventually increase when $\beta_{z,{\rm mid}}\gtrsim 10^5$ 
(Figure \ref{fig:btdep}). 
In order to take into account this effect we adopt the following 
prescription : 
\begin{equation}
\alpha = \alpha_{\rm fl} \times \max(1,\frac{\Sigma_{\rm up}(r)}{\Sigma(r)}),
\label{eq:so1}
\end{equation}
and
\begin{equation}
C_{\rm w} = C_{\rm w,fl} \times \max(1,\frac{\Sigma_{\rm up}(r)}{\Sigma(r)}),
\label{eq:so2}
\end{equation}
where 
$\Sigma_{\rm up}$ is the surface density at which $\alpha$ 
and $C_{\rm w}$ start to increase in Figure \ref{fig:btdep} 
($\beta_{z,{\rm mid}} \sim 10^5$).  
$\Sigma_{\rm up}$ is determined by the initial vertical magnetic flux. 
We model $\Sigma_{\rm up}(r) = \delta_{\rm up}\Sigma_{\rm init}(r)$;
$\alpha$ and $C_{\rm w}$ starts to increase
when the surface density decreases to $\delta_{\rm up}$ of the initial 
value. For simplicity, we assume a constant 
$\delta_{\rm up} = 0.01$ in this paper.

\begin{table}[h]
\tablenum{1}
\begin{tabular}{|c|c|c|c|}
\hline
Model & Disk Wind & net $B_z$ & Dead Zone \\
\hline
\hline
I & No  & Weak/No & No \\
\hline
II & Yes  & Weak/No & No \\
\hline
III & Yes  & Strong & No \\
\hline
IV & No & Weak/No & Yes \\
\hline
V & Yes & Weak/No & Yes\\
\hline
\end{tabular}
\label{tab:mdls}
\caption{Global models.}
\end{table}

We calculate 
the three cases for the disks without dead 
zones summarized in Table \ref{tab:mdls} (Models I -- III). 
In Models II and III 
we take into account the disk winds; Model II adopts the constant $\alpha$ 
\& $C_{\rm w}$ to model weak vertical magnetic fields and Model III prescribes 
Equations (\ref{eq:so1}) and (\ref{eq:so2}) to incorporate relatively strong 
vertical fields. 

\subsection{Disks with Dead Zones}
\label{sec:dzmd}
We apply the results of the local resistive MHD simulations (\S \ref{sec:dzls}
and Figure \ref{fig:dz2}) to the global model.
An essential point is that $\alpha$ and $C_{\rm w}$ can be determined by 
$r$ and $\Sigma$.    
Then, we use the following parameterization : 
\begin{equation}
\alpha = \alpha_{\rm fl} \frac{\Sigma_{\rm act}}{\Sigma},
\label{eq:dzal}
\end{equation}
where $\Sigma_{\rm act}$ is the sum of the column 
density of active regions near upper and lower disk surfaces which is 
modeled as
\begin{equation}
\Sigma_{\rm act} = \min\left(\Sigma_{\rm CR} + \Sigma_{\rm X}\left(
\frac{r}{1{\rm AU}}\right)^{-2},\Sigma\right),
\end{equation}
where $\Sigma_{\rm CR}$ is the column density with sufficient ionization 
by cosmic rays, and $\Sigma_{\rm X}$ is the column density with sufficient 
ionization by X-rays normalized at 1 AU. 
$\Sigma_{\rm X}$ has dependence on $r^{-2}$ to take into account 
the geometrical dilution of the X-ray flux from a central star, 
while $\Sigma_{\rm CR}$  
is constant because cosmic rays are diffusely distributed.  
To reproduce the results of the local simulations, we adopt $\Sigma_{\rm CR} 
= 12$ g cm$^{-2}$ and $\Sigma_{\rm X} = 25$ g cm$^{-2}$. 
The mass flux of disk winds is also reduced in accordance with $\alpha$, 
but we set a lower limit to match the local simulation results: 
\begin{equation}
\eqnum{S18}
C_{\rm w} = C_{\rm w,min} + (C_{\rm w,fl} - C_{\rm w,min})
\frac{\alpha}{\alpha_{\rm fl}}
\label{eq:dzcw}
\end{equation}
where we use $C_{\rm w,min} = 0.45\times C_{\rm w,fl}$.

The solid lines in Figure \ref{fig:dz2} represent Equations (\ref{eq:dzal}) 
and (\ref{eq:dzcw}). In the figure we use $\alpha_{\rm fl}=0.011$ and 
$C_{\rm w,fl} =7.7\times 10^{-5}$ for the floor values of the ideal MHD 
simulations with net vertical field of $\beta_{z,{\rm mid}}=10^6$ 
(Figure \ref{fig:btdep}).
For the global disk calculation 
we use the original 
values, $\alpha_{\rm fl}=8\times 10^{-3}$ and $C_{\rm w,fl}=2\times 10^{-5}$.

\section{Results of Global Models}
\label{sec:res}

\subsection{Evolution of Gas Disks}
\begin{figure}
\figurenum{9} 
\epsscale{1.1}
\begin{center}
\plotone{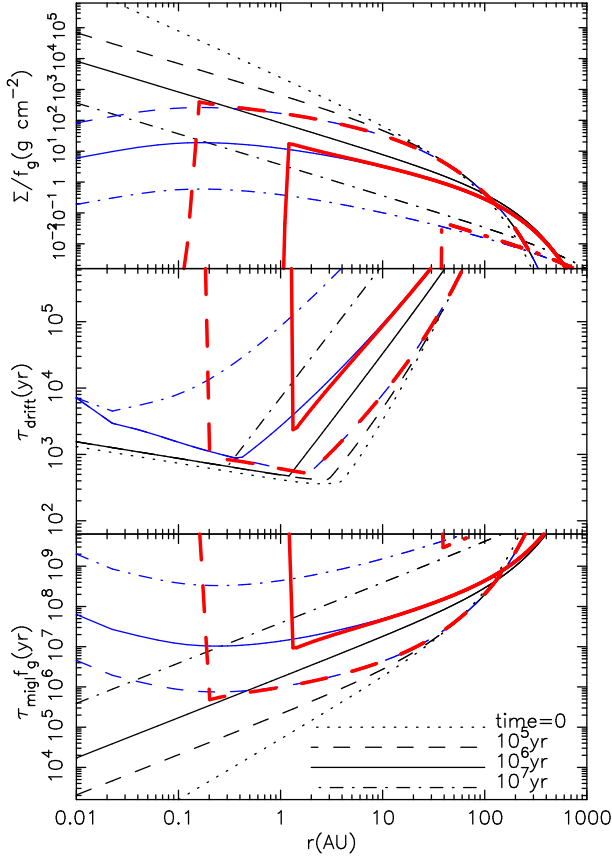}
\end{center}
\caption{The results of the no dead zone cases (Models I -- III). 
From top to bottom, time evolution of disk surface density, $\Sigma$, 
timescale, $\tau_{\rm drift}$, of inward drift of a m-size boulder, 
and timescale, $\tau_{\rm mig,I}$, of type I migration of an Earth-mass 
planet are shown. 
The black thin lines are the results without disk winds (Model I); 
the blue thin lines are the results with disk winds for weak/no vertical 
magnetic fields (Model II); 
the red thick lines the results with disk winds for relatively strong net 
vertical fields 
(Model III). 
The dotted lines are the initial values. The dashed, solid, and dot-dashed 
lines are the results at $10^5$, $10^6$, and $10^7$ yrs. 
Note that $\Sigma$ and $\tau_{\rm mig,I}$ can be scaled by $f_{\rm g}$. 
For example, the case with $f_{\rm g}=2$ gives twice larger $\Sigma$ and 
smaller $\tau_{\rm mig,I}$ than the case with $f_{\rm g}=1$. 
The results of $\tau_{\rm drift}$ are for $f_{\rm g} = 1$, because this 
scaling cannot be applied to $\tau_{\rm drift}$.
The bending points of $\tau_{\rm drift}$ correspond to the change of 
the regimes of drag force. 
The inside region corresponds to the Stoked regime where the dust size is 
larger than the mean free path of a gas particle, and the outside region 
corresponds to the Epstein regime where the dust size is smaller that 
the mean free path of a gas particle. 
}
\label{fig:sgev1}
\end{figure}

The top panel of Figure \ref{fig:sgev1} shows the evolution of the surface 
densities of the no dead zone cases (Models I -- III). 
The result of the no wind case (Model I; black thin lines) 
follows a self-similar solution of $\Sigma \propto 1/r$ in the inner
region with an exponential cutoff in the outer region \citep{lp74}.
On the other hand, the disk wind cases (Models II and III; blue thin and 
red thick lines) shows faster decreases of $\Sigma$ in the inner regions 
owing to the contribution from the disk winds in addition to the accretion. 
The figure clearly shows that the dispersal of the gas disk takes place in 
an inside-out manner, as discussed above (Equation \ref{eq:msfx}). 

The case with relatively strong vertical net magnetic flux (red lines) 
shows an expanding inner hole because 
the surface density decreases to reach $\delta_{\rm up}(=0.01)$ of the initial 
values faster in the inner regions and $\alpha$ and 
$C_{\rm w}$ start to increase from inner to outer locations.   
Although the quantitative properties of evolving inner holes depends 
on the adopted parameters (see \S \ref{sec:ndzls}), the observed properties 
of transitional disks with inner holes \citep{cal05,esp08,hug09} may be 
explained by the mechanism presented here. 

\begin{figure}
\figurenum{10} 
\epsscale{1.1}
\begin{center}
\plotone{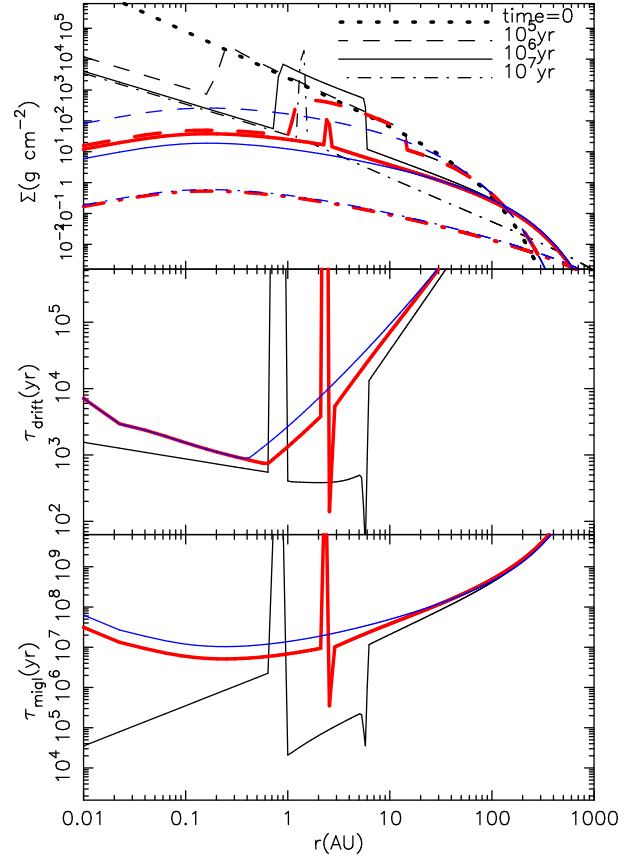}
\end{center}
\caption{Comparison of the dead zone cases with the no-dead zone case. 
The black, red, and blue lines are the results of Models IV (dead zone / 
no disk wind), V (dead zone / disk wind), and II (no dead zone / disk wind), 
respectively. 
The top panel compares the time evolution of surface density, $\Sigma$. 
The dotted, dashed, solid, and dot-dashed lines are the results at $t=0$, 
$10^5$, $10^6$, and $10^7$ yrs. The middle panel compares the inward drift 
timescales, $\tau_{\rm drift}$, of a m-size boulder at $t=10^6$ yr. 
The bottom panel compares the timescale, $\tau_{\rm mig,I}$, of type I migration 
of an Earth-mass planet at $t=10^6$ yr. 
}
\label{fig:sgev3}
\end{figure}

The top panel of Figure \ref{fig:sgev3} shows the evolution of the 
surface densities of the dead zone cases. 
The results of Models IV \& V exhibit density enhancements, 
which correspond to the dead zones. 
Because the surface densities in these regions are high, the X-rays and 
cosmic rays cannot penetrate to the midplanes. The dead zones form 
around the midplanes, and $\alpha$ and $C_{\rm w}$ become smaller. Smaller 
$\alpha$ leads to slower mass accretion\footnote{The disk expands around the
outer edge of the dead zone, while the mass accretes in the rest of the 
dead zone region.}, and then, the mass accumulates around the dead zones, 
which is observed as the density enhancements. 
Without disk winds, the dead zone exists until $10^7$ yr (Model IV; black 
lines).  

On the other hand, when taking into account the disk winds, the dead zone 
almost disappears at $10^6$ yr (Model V; red lines) because 
the surface density decreases faster owing to the disk winds, which 
further leads to the effective penetration of the ionizing 
X-rays and cosmic rays to the midplane. 
After the dead zone disappears, the disk evolution follows the 
no dead zone case. The surface density structure in later times 
(e.g. at $10^7$ yr in the figure) is very similar to that of the no dead  
zone case (Model II; blue lines). 
We can conclude that the MRI-driven disk winds play an essential role in 
the dispersal of protoplanetary gas disks even though the dead zone forms.

\subsection{Dynamics of Boulders and Planetesimal Formation}

\begin{figure}
\figurenum{11} 
\epsscale{1.1}
\begin{center}
\plotone{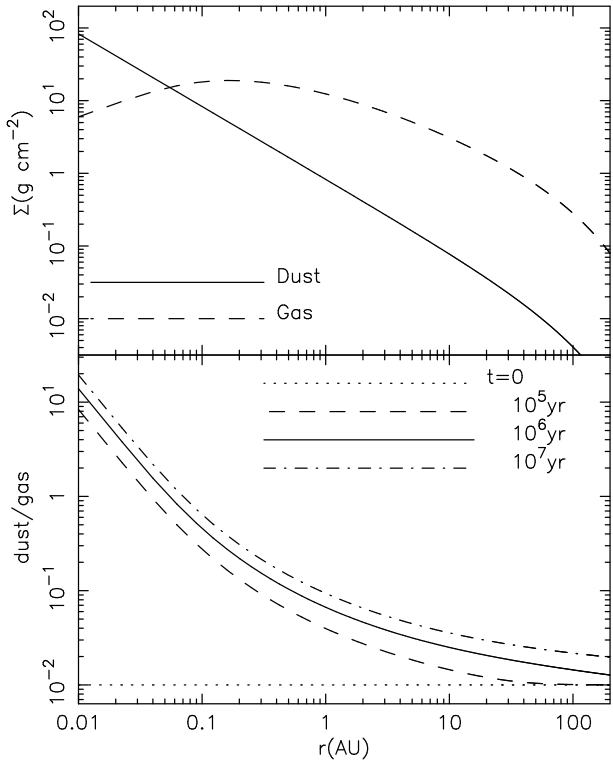}
\end{center}
\caption{{\it Upper}: Surface density of dust (solid) and gas (dashed) 
components at $t=10^6$ yr. {\it Lower}: Dust-to-gas ratios at $t=10^5$ yr
(dashed), $10^6$ yr (solid), and $10^7$ yr (dot-dashed). The dotted line 
is the initial condition (dust-to-gas ratio $=0.01$).} 
\label{fig:d2g}
\end{figure}

In addition to the dynamical evaporation of protoplanetary gas disks, 
MRI-driven disk winds affect the planet formation at various stages. 
At an early stage the rapid infall of boulders to a central star 
\citep{wei77}, 
which hinders the growth of solids to larger bodies by aggregation, 
is a severe problem.
The solid component rotates with the Keplerian velocity as a result of 
the force balance between the gravity due to a central star and the centrifugal
force. On the other hand, the gas rotates with sub-Keplerian velocity 
by the contribution from outward pressure gradient force. 
Then, the rotation of the solid component is slightly slowed down because 
of the head-wind from the gas, 
and drifts inward. 
Under the typical MMSN condition, the infalling timescale of $\sim$ 
meter-sized boulders is $\sim 100 - 1000$ years at 1 AU, which is too rapid 
to form planetesimals ($\sim$ kilometer size) in a turbulent gas disk. 

However, our calculations show that the surface density is increasing with 
$r$ in the inner region (the top panels of Figures \ref{fig:sgev1} and 
\ref{fig:sgev3}) . 
The inward drift rate becomes smaller than previously discussed. 
The middle panels of Figures \ref{fig:sgev1} and \ref{fig:sgev3} compares 
the inward drift timescales \citep{wei77},   
\begin{equation}
\tau_{\rm drift} = r\left(-\frac{1}{\Omega\rho}\frac{dp}{dr} 
\frac{t_{\rm s}\Omega}{1+(t_{\rm s}\Omega)^2}\right)^{-1}, 
\end{equation}
where $t_{\rm s}$ is stopping time of solid material in a gas disk. 
Here we consider a one meter-size spherical boulder for $t_{\rm s}$, 
and the pressure-gradient force ($dp/dr<0$ for sub-Keplerian rotating 
gas disks) is estimated at the midplanes of disks. 
As expected, $\tau_{\rm drift}$ becomes longer when the disk wind is 
taken into account (Figure \ref{fig:sgev1}), which 
is more favorable for the formation of planetesimals.
The tendency is the same for the dead zone cases as well 
(Figure \ref{fig:sgev3}), while $\tau_{\rm drift}$'s show 
complicated behaviors at the edges of the dead zones reflecting the sharp 
density gradients.

The dispersal of the gas component directly leads to the increase of 
a dust-to-gas ratio, which is also important in the context of gravitational 
instability of dust particles. 
\citet{sek98} investigated the turbulence due to shear motions between gas 
and dust. He found that when a significant fraction of the gas component 
is dispersed with dusts left, the shear-induced turbulence is reduced and 
dusts become gravitationally unstable, which possibly leads to 
the formation of planetesimals \citep[see also][for related works]{ys02,mi06}. 
\citet{joh07} also proposed that streaming instability triggers the direct 
formation of large planetesimals or dwarf planets when the dust-to-gas ratio 
increases to an order of unity.  
The disk winds disperse the gas component selectively and 
 raise the dust-to-gas ratio in an inside-out manner, 
 which activates streaming instability from inner regions. 

To illustrate the increase of the dust-to-gas ratio, we calculate the 
time-evolution of dust surface density by using the same global model. 
Here, we assume that dust particles follow the only viscous accretion 
with gas near the midplane and do not escape with disk winds; 
we solve Equation (\ref{eq:sgmev}) by setting $(\rho v_z)_{\rm w}=0$ for 
the dust component.
This assumption is reasonable if dust particles are well-coupled with 
gas near the midplane and decoupled in the upper regions. 
Nondimensional stopping time, $\Omega t_s$, is a good indicator 
which measures the coupling between dust and gas; if $\Omega t_s<1$, 
dusts are well-coupled with gas, and vice versa. 
When one takes dust particles with size of millimeter at 1 AU of the 
MMSN with $f_g=1$ (Equation \ref{eq:sgmmsn}) as an example, 
$\Omega t_s\approx 2\times 10^{-4}$ at the midplane. 
Since stopping time is inversely proportional to 
gas density, $\Omega t_s$ exceeds unity at $z\approx \pm 3H$ where 
$\rho/\rho_{\rm mid}\approx 2\times 10^{-4}$ (Figure \ref{fig:dz1}).
The assumption is reasonable for moderately small dusts (sub-millimeter -- 
meter at 1 AU of the MMSN). Further smaller dusts will be lift up with disk 
winds, while larger solid particles are decoupled with gas even at the 
midplane.

We initially impose a uniform dust-to-gas ratio $=0.01$ and follow 
the evolution of both gas and dust with using the parameters of Model II 
of Table \ref{tab:mdls}.  
The upper panel of Figure \ref{fig:d2g} presents the surface densities of 
dust (solid) and gas 
(dashed) at $t=10^6$ yr, which shows that the dust surface density is 
larger than the gas surface density in the inner region, $r\lesssim 0.1$ AU.
The lower panel shows that the dust-to-gas ratio gradually increases from 
the inner region.   



\subsection{Type I Migration}

\begin{figure}
\figurenum{12} 
\epsscale{1.}
\begin{center}
\plotone{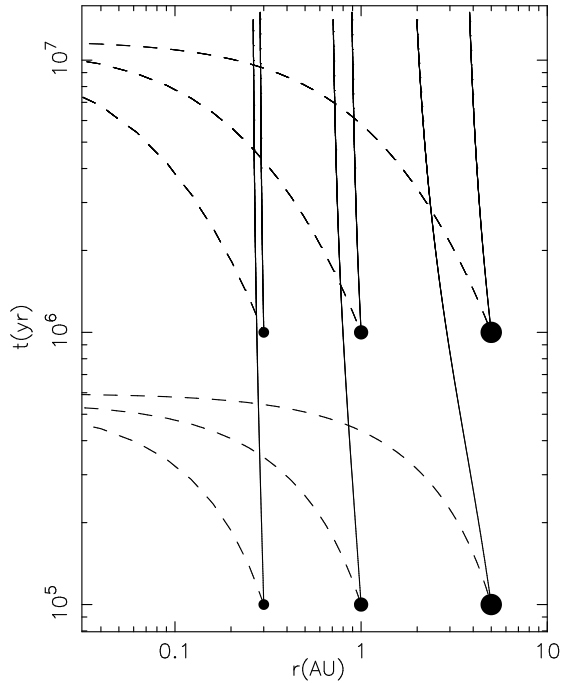}
\end{center}
\caption{Orbital evolution of planets by type I migration. 
The horizontal axis shows radial distance and the vertical axis shows 
time elapsed from the starting point of the disk calculation.  
The solid lines are the results with disk winds (Model II) and the dashed 
lines are the results without disk wind (Model I). 
The circles indicate the initial locations of newly formed planets : 
At $t=10^5$ yr and $10^6$ yr we put 
planets with $0.3M_{\oplus}$ at 0.3 AU, planets with $1 M_{\oplus}$ at 1 AU, 
and planets with $5M_{\oplus}$ at 5 AU.  }
\label{fig:plmg}
\end{figure}

After the formation of planets, the gas component in a disk also 
 plays a role in the evolution of the planetary system. 
A lower-mass planet,
which cannot create a gap in a gas disk, 
resonantly interacts with the gas component of a disk through gravitational 
torque \citep{wad97}.   
As a result planets generally migrate inward 
 with timescale \citep{ttw02}, 
\begin{equation}
\tau_{\rm mig,I}(r) \approx 5\times 10^4 {\rm yr}
\left(\frac{4.35}{2.7+1.1s}\right)
\left(\frac{\Sigma(r)}{\Sigma_0}\right)^{-1}
\left(\frac{M}{M_{\oplus}}\right)^{-1}
\label{eq:tpImg}
\end{equation}
where $s$ is the local gradient of $\Sigma \propto r^{-s}$, $M$ is planet mass, 
$M_{\oplus}$ is the Earth mass, and we assume the MMSN around a central star 
with the solar mass. 
This equation indicates that the migration is faster in a more 
massive (larger $\Sigma$) gas disk because of the larger torque on a planet. 
In typical situations, $\tau_{\rm mig,I}$ is shorter than the lifetimes of 
 protoplanetary gas disks; 
 newly formed terrestrial planets and cores of gas-giant planets quickly fall 
 into a central star. 

However, we have shown that the gas density in the inner regions quickly 
decreases by the disk winds, and the gradient of surface density 
becomes positive in the inner region. 
Consequently, 
$\tau_{\rm mig,I}$ becomes considerably longer than previously considered. 
The bottom panels of Figures \ref{fig:sgev1} and \ref{fig:sgev3} show 
$\tau_{\rm mig,I}$ for an Earth-mass planet around a solar-mass star. 
The figures illustrate that the disk winds greatly 
suppress the migrations in the inner region both in no dead zone and dead zone 
cases. 
Similarly to $\tau_{\rm drift}$, $\tau_{\rm mig,I}$'s of the dead zone cases 
show complicated behaviors at the edges of the dead zones \citep{mpt07,kat09}.

To illustrate the suppression of type I migration more clearly, 
we calculate the migrations of planets with the 
evolution of protoplanetary disks 
by using migration speed, $r/\tau_{\rm mig,I}$ (Figure \ref{fig:plmg}). 
In the no disk wind case (Model I) all the planets infall to a central star. 
On the other hand, in the disk wind case (Model II) the migrations are slow   
and all the planets survive.  
When the disk winds are considered, type I migration becomes unimportant 
especially at later times, $t\gtrsim 10^5 - 10^6$ yr, under the typical MMSN 
condition. 

\vspace{2cm}
\section{Discussions}
\label{sec:dis}

\subsection{Energetics}
\label{sec:eng}
In this paper we have applied results of the local simulations to 
the global models. 
We use the mass flux of the disk winds at the upper and lower boundaries, 
$z=\pm 4H$, of the simulation box. 
Since the wind velocities at the boundaries are still smaller than the escape 
speed from a central star, we should carefully examine whether the 
disk winds really escape from the disks. 
In this subsection, we examine the energetics of the disk winds to see 
whether the accretion energy can potentially drive the disk winds that 
can escape from the gravity of a central star. 
 
The starting point here is an equation for the conservation of total energy :
$$
\hspace{-2cm}
\frac{\partial}{\partial t}\left[\rho\left(\frac{1}{2}v^2 
+ \frac{1}{\gamma-1}\frac{p}{\rho} - \frac{G M_{\star}}{r}\right) 
+ \frac{B^2}{8\pi}\right]
$$
$$
\hspace{-0.5cm}
+ 
\mbf{\nabla\cdot}
\left[\left\{\rho \mbf{v}\left(\frac{1}{2}v^2 
+ \frac{\gamma}{\gamma-1}\frac{p}{\rho}
-\frac{G M_{\star}}{r}\right) \right.\right.
$$
\begin{equation}
\left.\left. + \mbf{v} 
\frac{B^2}{4\pi} - \frac{\mbf{B}}{4\pi}(\mbf{B}\cdot\mbf{v})\right\}\right] 
= - q_{\rm loss},
\label{eq:eng3d}
\end{equation}
where $q_{\rm loss}$ is energy loss which is modeled below and here 
we consider the only $r$ derivative.
We integrate Equation (\ref{eq:eng3d}) with $\int dz$ 
by neglecting the small terms except the anisotropic stress, 
$\delta v_{\phi} v_r - B_{\phi}B_r/4\pi\rho = \alpha c_s^2$, whereas
we separate $v_{\phi}$ into Keplerian rotation plus small 
perturbation, $v_{\phi}= r\Omega + \delta v_{\phi}$. 
We also assume that accretion velocity and sound speed 
($c_s=\sqrt{\gamma p/\rho}$) are smaller 
than rotation velocity,
$v_r, c_s\ll r\Omega$.
Then, the total energy of a ring at $r$ changes according to 
\begin{equation}
\frac{\partial}{\partial t}\left[-\Sigma\frac{r^2\Omega^2}{2}
\right] + \frac{1}{r}\frac{\partial}{\partial r}\left[r\Omega 
\frac{\partial}{\partial r}(r^2 \Sigma \alpha c_s^2) +r^2\Sigma 
\Omega \alpha c_s^2\right] = - Q_{\rm loss} ,
\label{eq:tteng}
\end{equation}
where $Q_{\rm loss} = \int q_{\rm loss} dz$ and we have used 
$\frac {G M_{\star}}{r^2} = r \Omega^2$ and $r\Sigma v_r = \frac{2}{r\Omega} 
\frac{\partial}{\partial r}(r^2 \Sigma \alpha c_{s}^2)$. 
The spatial derivative term on the left-hand side represents the energy 
liberated by mass accretion and viscous heating.

We investigate whether this liberated gravitational energy  
is sufficient to drive disk winds. For simplicity, we only consider the 
kinetic energy of disk winds for the loss term and neglect additional 
effects such as heating by UV/X-ray radiation (energy input), acceleration 
by stellar winds (momentum input), 
and radiative cooling (energy loss).   
Since we consider the disk winds from a Keplerian rotating disk, we can 
write $Q_{\rm loss}=\frac{1}{2}\rho v_{z}(v_z^2 + r^2 \Omega^2)$.
Disk winds can escape to infinity if $v_z$ exceeds the escape speed, 
$v_{\rm esc} = \sqrt{2}r\Omega$; namely if the condition, 
$$
\frac{\partial}{\partial t}\left[\Sigma\frac{r^2\Omega^2}{2}
\right] - \frac{1}{r}\frac{\partial}{\partial r}\left[r\Omega 
\frac{\partial}{\partial r}(r^2 \Sigma \alpha c_s^2) +r^2\Sigma 
\Omega \alpha c_s^2\right] 
$$
\begin{equation}
- \frac{3}{2}(\rho v_z)_{\rm w} r^2 \Omega^2 \ge 0
\label{eq:engcd}
\end{equation} 
is satisfied, disk winds can be driven solely by the liberated 
gravitational energy of accretion, because the left-hand side of 
Equation (\ref{eq:engcd}) is the energy flux of disk winds at infinity. 
The time derivative term in Equation (\ref{eq:engcd}) can be eliminated 
by using Equation (\ref{eq:sgmev}). Then, using the relation of Keplerian 
rotation, $\Omega \propto r^{-3/2}$, Equation (\ref{eq:engcd}) is rewritten 
as
\begin{equation}
\frac{3}{2}\Omega \Sigma \alpha c_s^2 \ge 2r^2 \Omega^2 (\rho v_z)_{\rm w},
\end{equation}
or more specificly, 
\begin{equation}
r\ge \frac{8}{9\pi}\frac{C_{\rm w}^2 (r_0 \Omega_0)^4}{\alpha^2 c_{s,0}^4}r_0 
\equiv r_{\rm dw}, 
\end{equation}
where we are using Equation (\ref{eq:ndms}) and the relation of the MMSN 
($c_s\propto r^{-1/4}$) with Keplerian rotation. 
Substituting the standard values, $\alpha=8\times 10^{-3}$ and $C_{\rm w} 
=2\times 10^{-5}$, in the global model we have $r_{\rm dw} = 1.4$ AU.

The disk winds in the inner region, $r<r_{\rm dw}$, do not have sufficient 
energy to escape from a disk by accretion. 
The fates of these wind materials 
are (i) accreting directly to a central star 
if the angular momentum is removed, (ii) blown away by the stellar winds 
(see \S \ref{sec:stw}), or (iii) returning back to the disk 
(see Figure \ref{fig:ctn} for the schematic picture).
If the most of the wind gas follows the processes (i) or (ii), 
our calculations in \S \ref{sec:res} give the correct surface gas densities. 
On the other hand, if the process 
(iii) is dominant, we overestimate the escaping mass flux of the disk winds. 
We further discuss the escape of the disk winds in \S \ref{sec:esc}.

\subsection{Global Modeling with Energetics}
\begin{figure}
\figurenum{13} 
\epsscale{1.}
\begin{center}
\plotone{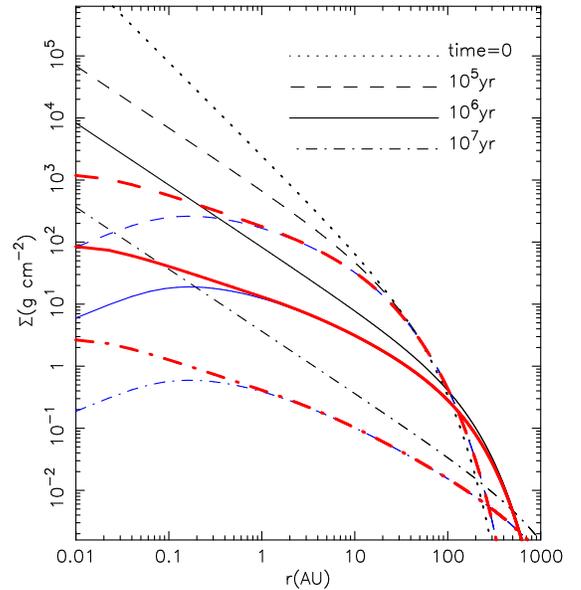}
\end{center}
\caption{Time evolution of disk surface density. 
The black thin lines are the results without disk winds (Model I); 
the blue thin lines are the results with the disk winds of the constant 
mass flux, $C_{\rm w}=C_{\rm w,fl}$ (Model II); the red thick lines the 
results with the disk winds which take into account the energetics limiter 
for $C_{\rm  w}$ (Equation \ref{eq:clmt}).  
The dotted lines are the initial values. The dashed, solid, and 
dot-dashed lines are the results at $10^5, 10^6$, and $10^7$yrs. 
}
\label{fig:sgev2}
\end{figure}
\vspace{1cm}
We can take into account the energetics argument (\S \ref{sec:eng}) 
in the global model.  
We adopt the following limiter for disk wind mass flux, $C_{\rm w}$ : 
\begin{equation}
C_{\rm w} = \min(C_{\rm w,fl}, C_{\rm w,eng}), 
\label{eq:clmt}
\end{equation} 
where $C_{\rm w,eng}$ corresponds to the mass flux that gives the left-hand 
side of Equation (\ref{eq:engcd}) equal to zero, namely 
$$
\hspace{-4cm}C_{\rm w,eng} = \left\{\frac{\partial}{\partial t}
\left[\Sigma\frac{r^2\Omega^2}{2}\right] \right .
$$
\begin{equation}
\left. - \frac{1}{r}\frac{\partial}{\partial r}\left[r\Omega 
\frac{\partial}{\partial r}(r^2 \Sigma \alpha c_s^2) +r^2\Sigma 
\Omega \alpha c_s^2\right] \right\}
\left[\frac{3}{2} r^2 \Omega^2 \rho_{\rm mid} c_s\right]^{-1}. 
\end{equation} 
Equation (\ref{eq:clmt}) reduces the mass flux in $r<r_{\rm dw}$ to the value 
available from the liberated gravitational energy by accretion. 
In $r\ge r_{\rm dw}$, Equation 
(\ref{eq:clmt}) gives $C_{\rm w} = C_{\rm w,fl}$.

Figure \ref{fig:sgev2} displays the results which take into account the 
limiter for $C_{\rm w}$ (thick red lines) in Model II, in comparison with 
Model I (no disk wind; black thin lines) and Model II 
(constant $C_{\rm w}$; blue thin lines). 
As expected, the difference 
between the red and blue lines appears only in $r<r_{\rm dw}$($=1.4$ AU), the 
surface densities in the outer region are the same. 
Although the surface density 
in the inner region becomes larger in the case with the $C_{\rm w}$ limiter, 
it is still much lower than the surface 
density of the no wind case. Then, the disk winds still play an essential 
role in the dispersal of protoplanetary disks even though taking into account 
the energetics of the disk winds.  

\subsection{Mass Loss Rate}
\begin{figure}
\figurenum{14} 
\epsscale{1.}
\begin{center}
\plotone{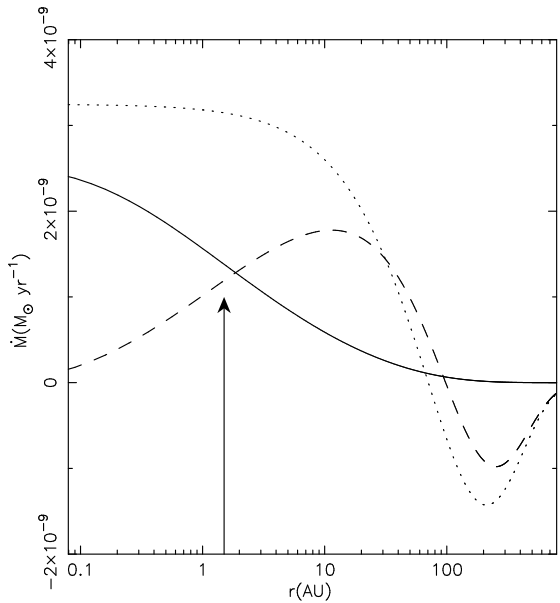}
\end{center}
\caption{Structure of mass loss/accretion rates of 
Model II in the main paper at $t=10^6$ yr. 
The solid line is 
the mass loss by the disk winds (Equation \ref{eq:mdwd}) and the dashed 
line is the mass accretion rate (Equation \ref{eq:macc}) of the model  
taking into account the disk winds. 
The dotted line is the mass accretion rate of the model without disk winds 
for comparison. The arrow indicates the location of $r_{\rm dw}$. } 
\label{fig:engacc}
\end{figure}

Figure \ref{fig:engacc} shows the mass accretion rate, 
$\dot{M}_r$, and the mass loss rate of disk winds, $\dot{M}_z$, which are 
respectively defined as 
\begin{equation}
\dot{M}_r = - 2\pi r\Sigma v_r , 
\label{eq:macc}
\end{equation}
and
\begin{equation}
\dot{M}_z(r) = 2\pi \int_{r}^{r_{\rm out}} r' dr' (\rho v_z)_{\rm w}. 
\label{eq:mdwd}
\end{equation} 
In the case with disk winds, the mass accretion rate decreases for 
decreasing $r$ because the mass is lost by the disk winds. 
The total mass loss rate by the disk winds is 
$\dot{M}_z(r_{\rm in}) =2.5\times 10^{-9}$ $M_{\odot}$ yr$^{-1}$ and  
the mass loss rate from $r>r_{\rm dw}$ that can be accelerated to infinity 
by accretion energy is $\dot{M}_z(r_{\rm dw}) = 1.4\times 10^{-9}$ 
$M_{\odot}$ yr$^{-1}$; more than the half of the total wind mass loss 
can escape from the disk by the liberated gravitational energy. 
The mass loss by the disk winds is larger than the accretion rate. 
Assuming the truncation of the disk at 5-10 stellar radii ($\sim 0.1$ AU), 
$\dot{M}_r \approx 2\times 10^{-10}$ $M_{\odot}$ yr$^{-1}$, which can be 
regarded as the mass 
accretion rate through magnetic flux tubes directly connecting to the central 
star \citep{ken96}. 

\begin{figure}
\figurenum{15} 
\epsscale{1.1}
\begin{center}
\plotone{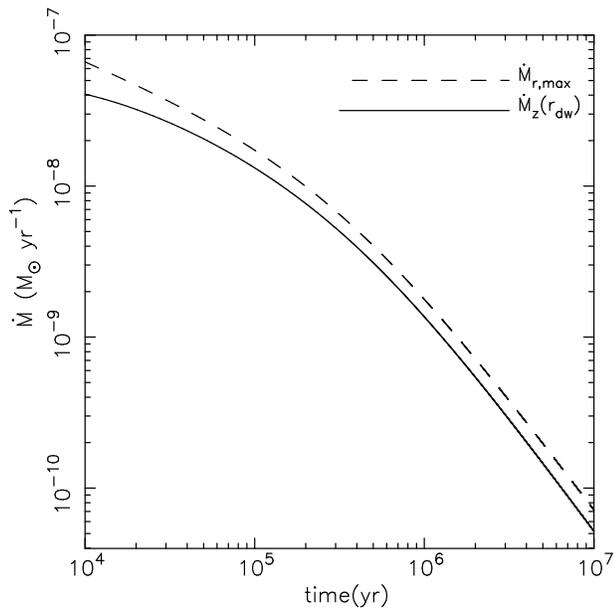}
\end{center}
\caption{Time evolution of the mass loss rate of the disk wind (solid) and 
accretion rate (dashed) of Model II. 
Here, the wind mass loss rate is the integration of $r>r_{\rm dw}$, 
$\dot{M}_z(r_{\rm dw})$, that can potentially escape from a central star 
gravity by accretion energy. The shown accretion rate is the maximum value 
at each time (e.g. for $t=10^6$ yrs in Figure \ref{fig:engacc}, the maximum 
accretion rate is obtained at $r\approx 15$ AU). } 
\label{fig:tmevmd}
\end{figure}

The obtained mass loss rate by the MRI-driven disk winds is larger than the 
mass loss rate, $\lesssim 10^{-10} - 10^{-9}$ $M_{\odot}$ yr$^{-1}$, 
predicted by the UV photoevaporation (e.g. Matsuyama et al.2003). 
Recently, photoevaporation by X-rays from a central star is also proposed. 
It is reported that the mass loss rate due to X-rays may be larger 
than that by the UV photoevaparation, where the calculated mass loss 
rates are rather uncertain 
\citep{ecd09,gdh09,owe10}. Significant difference of the disk wind mass loss 
from the photoevaporation processes by UV or X-rays is time evolution.  
The mass loss rate by the disk winds is correlated with the accretion rate 
(Figure \ref{fig:tmevmd}), because the energy source of the disk winds is 
the gravitational energy liberated by accretion. 
Therefore, the wind mass loss rate is larger at earlier times when the 
accretion rate is larger; the disk winds significantly 
contribute to the dispersal of the gas component of a disk from the beginning. 
As a result, an inner hole forms from the early epoch and its size is gradually 
expanding.
On the other hand, the UV photoevaporation mechanism is significant at the 
later times after the sufficient mass dissipates by accretion. 
Therefore, an inner hole is anticipated at later time when the evaporating 
mass flux becomes comparable to the mass accretion rate (Alexander et al.
2006). 
The X-ray photoevaporation, which may be effective from slightly 
earlier time, is also expected to give the similar trend to the UV 
photoevaporation \citep{gdh09}. 

\subsection{Escape of Disk Winds}
\label{sec:esc}
We further continue the 
discussions on the escape of disk winds. 

\subsubsection{Local Simulations with Larger Vertical Boxes}
\label{sec:lslvb}
In order to study the acceleration of the disk winds at higher altitudes, 
we perform the local 3D MHD simulations with larger vertical boxes. 
We here use the realistic vertical gravity, 
\begin{equation}
g_z = \frac{G M_{\star} z}{(r^2+z^2)^{3/2}} = \Omega^2 z
\frac{r^3}{(r^2+z^2)^{3/2}}, 
\label{eq:zgr}
\end{equation}
where $r$ is radial location from a central star. 
In the usual local simulations, 
the only leading term, $g_z\simeq \Omega^2 z$, is 
considered,
because $r$ does not appear explicitly and it can be treated more easily 
(Hawley et al.1995). On the other hand, the escape velocity is not defined 
at the expense of the simplification, and the vertical gravity is overestimated 
by a factor of $\frac{r^3}{(r^2+z^2)^{3/2}}$, which makes density 
at a high altitude unrealistically lower. 
To avoid these shortcomings
we use Equation (\ref{eq:zgr}) by explicitly setting $r$
(Table \ref{tab:lzb}). Note that $r=20H$ corresponds to the location 
of $r\approx 1$ AU for the MMSN.  
In these runs, we set the net vertical fields with 
$\beta_{{z,\rm mid}}=10^6$ at the midplanes. 

\begin{figure}
\figurenum{16} 
\epsscale{1.2}
\begin{center}
\plotone{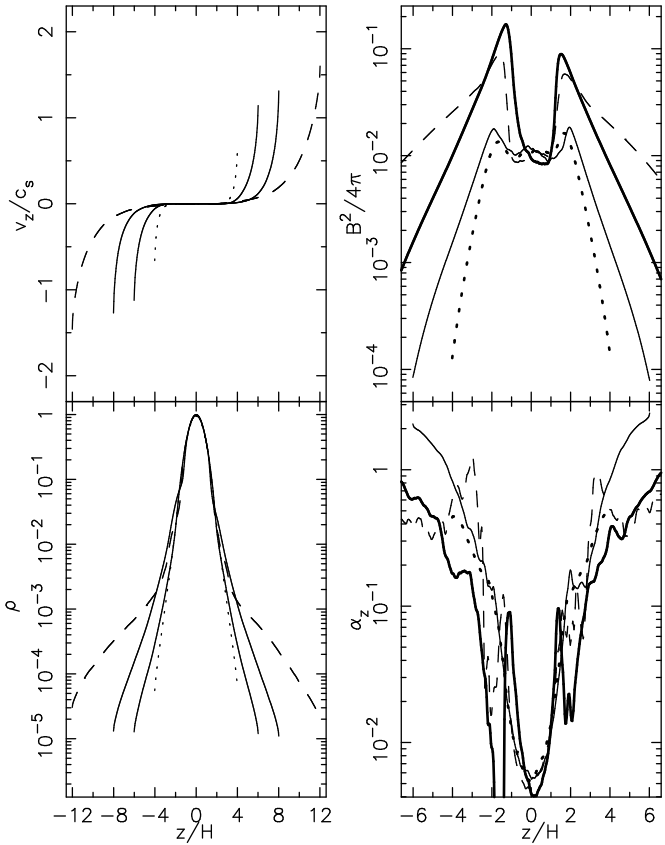}
\end{center}
\caption{Comparison of the disk wind structures with the different box sizes. 
On the left, we display the time-averages of 
vertical velocity and density. On the right we compare the time-averages of 
magnetic energies and $\alpha$ values only in the $-6H<z<6H$ region. 
The solid lines are the results with 
$r=20H$ and box sizes, $-6H < z < 6H$ (Model VI) and $-8H < z < 8H$ 
(Model VII). The dashed lines are the results with $r=10H$ and box size, 
$-12 H < z < 12 H$ (Model VIII). The dotted lines are the results 
of the reference case (Model II; $r\rightarrow \infty$ and 
$-4H < z < 4H$). On the right we use the thick lines for Model VII.}
\label{fig:lzb}
\end{figure}

\begin{table}[h]
\tablenum{2}
\begin{tabular}{|c|c|c|}
\hline
Model & $r$ & Box Size \\ 
\hline
\hline
II(Reference) & -- & $-4H < z < 4H$ \\ 
\hline
VI & 20 & $-6H < z < 6H$ \\ 
\hline
VII & 20 & $-8H < z < 8H$ \\ 
\hline
VIII & 10 & $-12H < z < 12H$ \\ 
\hline
\end{tabular}
\caption{Radial positions and vertical box sizes of the local simulations. }
\label{tab:lzb}
\end{table}

\begin{figure}
\figurenum{17} 
\epsscale{0.9}
\begin{center}
\plotone{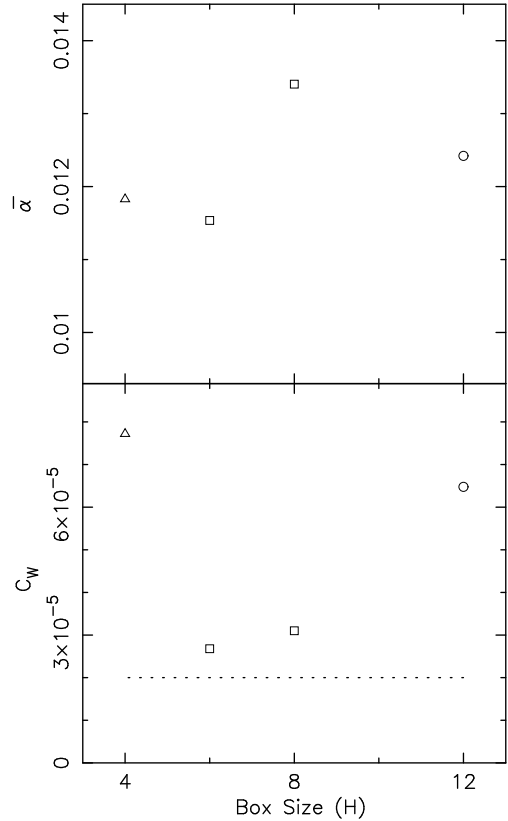}
\end{center}
\caption{The $\bar{\alpha}$ values (top) and the mass flux of the disk winds 
(bottom) for the different sizes of the 
vertical simulation boxes. The triangles are the results of Model II 
($r\rightarrow \infty$), the squares are the results with $r=20 H$ (Models 
VI \& VII), and the circles are the results with $r=10H$ (Model VIII). 
The dotted line is the level of 
$C_{\rm w,fl}$.
}
\label{fig:lzmf}
\end{figure}

Figure \ref{fig:lzb} compares the time averaged vertical structures of these 
cases (Table \ref{tab:lzb}). 
The top left panel shows that the onsets of the disk winds take place at higher 
altitudes in larger box cases; the results depend on the simulation box size. 
However, the mass flux of the disk winds ($\rho v_z$) do not show a monotonic 
behavior as presented in the bottom panel of Figure \ref{fig:lzmf}. 
By increasing the box size from $\pm 4H$ to $\pm 6H$, the mass flux decreases 
at first \footnote{Although we do not take into account $r$ in the 
case with the vertical box of $-4H < z < 4H$, this effect is negligible for 
sufficiently small $z$.}. 
However, increasing the box size from $\pm 6H$ to $\pm 8H$, the mass flux 
increases slightly. 
The largest box size case with the smaller gravity, $r=10H$, shows larger mass 
flux than these smaller box cases. 
The mass flux of the disk winds is bound by a lower limit and 
dose not becomes further smaller even though we use a larger vertical box. 

The dependence of the mass flux on the simulation box size can be explained by 
the saturation of the amplified magnetic fields (the top right panel of 
Figure \ref{fig:lzb}). 
When we increase the box size from $\pm 4H$ to $\pm 6H$, the escaping 
mass from the disk surfaces becomes smaller at first.  
Accordingly, the escaping magnetic flux of the toroidal ($y$) and radial 
($x$) components decrease, because the magnetic fields are frozen in the gas. 
The amplification of magnetic fields 
is balanced with the escape with the disk winds in addition to magnetic 
reconnections. 
In the larger box cases the escaping flux becomes smaller and the saturated 
level of magnetic fields increases. 
As a result of the larger magnetic pressure, the mass is lift up to higher 
locations 
(the bottom left panel of Figure \ref{fig:lzb}). 
The increase of the density inhibits further decrease of the mass flux of 
the disk winds, $\rho v_z$, in a self-regulated manner.      

In spite of the increase of the magnetic field strength for larger 
boxes, $\alpha_z$ does not increase so much (the lower right panel of 
Figure \ref{fig:lzb}). 
This is because the magnetic field of the 
large box cases around $z\approx 1.5H$ (the locations of the peaks) is 
dominated by the coherent toroidal component which does not contribute to 
the anisotropic Maxwell stress. Therefore, the integrated $\bar{\alpha} 
(= \int dz \rho \alpha_z/\int dz\rho )$, which directly determines the 
global mass accretion rate, does not depend on the vertical box size (the top 
panel of Figure \ref{fig:lzmf}).

The top left panel of Figure \ref{fig:lzb} shows that the average velocities 
of the disk winds do not still reach the escape speeds 
($=\sqrt{2}r\Omega = \sqrt{2}\left(\frac{r}{H}\right)H\Omega = 2
\left(\frac{r}{H}\right)c_s$). However, our conservative choice of the floor 
values, $C_{\rm w,fl}$, probably gives a reasonable estimate 
(the dotted line of Figure \ref{fig:lzmf}), 
first because the mass flux seems bounded by the lower limit as explained 
above, and second because there are a couple of mechanisms that favor 
the escape of the disk winds but are not included in this paper (see below).

\subsubsection{Magnetocentrifugal Winds}
Our local simulations do not take into 
account the acceleration (momentum input) of the disk winds by centrifugal 
force with global magnetic fields. 
If poloidal ($r-z$ components) magnetic field lines sufficiently incline with 
respect to an accretion disk, the gas can flow out along 
with the field lines by the centrifugal force \citep{bp82,ks98}. 
Such global magnetic fields will be common in protoplanetary disks as a 
result of the contraction of cold molecular clouds with 
interstellar magnetic fields. 
Therefore, our local simulations probably underestimate the momentum input to 
the disk winds. 


\subsubsection{Stellar Winds}
\label{sec:stw}

\begin{figure}
\figurenum{18} 
\epsscale{1.}
\begin{center}
\plotone{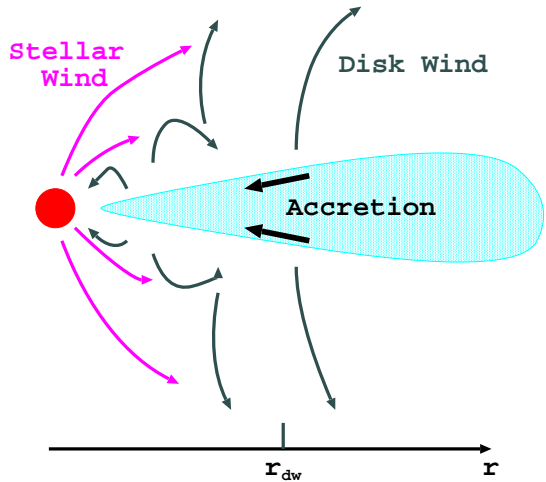}
\end{center}
\caption{Schematic picture of dispersal of a protoplanetary disk by disk 
winds. The wind material in the outer region, $r>r_{\rm dw}$, can stream 
out solely by the gravitational energy by accretion.
On the other hand, the wind material in the inner location
cannot escape by itself from the gravity of a central star. A fraction of 
the winds directly accrete to the central star after lift up. 
A fraction of the disk winds may be accelerated by the dynamical pressure 
of the stellar winds.  If the stellar wind flux is not strong enough, 
the disk wind material returns back to the disk after transported outward.  
}
\label{fig:ctn}
\end{figure}


So far, we have focused on the disk winds driven by MHD turbulence. 
Central T Tauri stars are also expected to drive stellar winds by 
the mass accretion from disks \citep{hir97,mp05} and 
the surface convections \citep{cra08}. 
Observations of T Tauri stars show that the outflow rates range from 
$10^{-11}$ to $10^{-8}$ $M_{\odot}$ yr$^{-1}$, which is typically 0.01 - 1 
times of the mass accretion rates \citep{wh04}. 
Although it is difficult to determine the exact launching locations of the 
observed outflows, some fractions are expected to come from central stars 
\citep{ed06}. 

At present the stellar wind is regarded to give a minor effect to the 
dispersal of the gas component of protoplanetary disks \citep[e.g.][]{shu93}
because the stellar winds almost slide along the disk surfaces \citep{mat09}.
However, if disk materials are lifted up as we have shown so far, 
the ram pressure of stellar winds can directly push away the lifted up 
materials because of the large elevation angle.  

We discuss the role of stellar winds from a simple momentum balance. 
Let us consider a situation, in which the lifted up gas by disk winds 
floats above a disk and the radial stellar winds hit the floating 
gas. If we assume that both lifted up gas and stellar wind gas move together 
radially outward after the hitting, we can estimate the radial velocity of the 
moving gas, $v_{\rm cmb}$, from the momentum balance as 
\begin{equation}
(\dot{M}_{z,{\rm in}} + W \dot{M_\star})v_{\rm cmb} = W \dot{M_\star} v_{\star}, 
\label{eq:stw}
\end{equation}
where $\dot{M}_{z,{\rm in}}$ is the rate of the mass that is supplied from 
disk winds but float above a disk without sufficient energy, $W$ is the 
fraction of the solid angle obscured by the lifted up gas by the disk winds, 
and $\dot{M}_{\star}$ and $v_{\star} (\approx 200-400\; {\rm km\;s^{-1}})$ are 
the mass loss rate and velocity of the stellar winds. 
As a typical example, we consider the result of Model II at 
$t=10^6$ yr (Figure \ref{fig:engacc}). Then, $\dot{M}_{z,{\rm in}}=1.1\times 
10^{-9}$ $M_{\odot}$ yr$^{-1}$; we assume the lifted up gas fill up 
to the height of $r/2$, which gives $W = \int_{\pi/2}^{\cot^{-1}(1/2)}
d\cos\theta \approx 0.45$. 
The gas that is lifted up by the disk winds but does not have the sufficient 
energy will distribute in $r<r_{\rm dw}(=1.4\;{\rm AU})$. Here, we compare 
$v_{\rm cmb}$ with the escape velocity at 1 AU, 
$v_{\rm esc, 0}\approx 42$ km s$^{-1}$ as a typical condition. 
Substituting these values into Equation (\ref{eq:stw}), 
if $\dot{M}_{\star}>4\times 10^{-10}$ $M_{\odot}$ yr$^{-1}$,  
$v_{\rm cmb} > v_{\rm esc,0}$ is satisfied and the lifted up material can be 
blown away by the stellar winds.


If $\dot{M}_{\star}$ is smaller than this value, the stellar winds can 
blow away the disk wind material at sufficiently high altitudes where 
the density is low. 
The disk wind material at lower heights will move outward after hit by 
the stellar winds but again return back to the disk without sufficient energy. 
When the returning location becomes $r > r_{\rm dw}$, the gas finally 
flow out by the disk winds.  
(see Figure \ref{fig:ctn} for the schematic picture).

\vspace{1cm}
\section{Summary}
In this paper, we have shown that the MRI-driven protoplanetary disk winds 
disperse the gas component of disks from the inside out. 
If net vertical magnetic fields with moderate strength exist, the disk winds 
and accretion switch-on from the inner locations, which forms an 
expanding inner hole. This mechanism naturally explains observed transitional 
disks with inner holes. 
Model calculations that incorporate UV or X-ray
photoevaporation with 
accretion also expects an inner hole at the later stage after the significant 
fraction of the gas disappears \citep{alx06,gdh09}. 
The main difference of our mechanism from the photoevaporation processes 
is that the MRI-driven disk winds expect an inner hole from the early 
times and its size gradually grows from $<0.1$ AU to several tens AU during the 
evolution of $\sim 10^7$ years.

Future high resolution observations 
by ALMA will be able to resolve inner holes with $\sim $ a few AU at 
distance of 100 pc. 
We hope that observations of protoplanetary disks with various epochs will 
reveal the time-evolution of inner holes.

The dead zone does not affect the disk winds so much and the effect is only 
limited at the early epoch ($\lesssim 10^6$ years) of the disk evolution, 
because the MRI-driven disk winds are driven from the surface regions with 
sufficient ionization degree. 
Even though a large dead zone forms around the midplane, the disk winds 
are also driven intermittently with quasi-periodic cycles of 5-10 rotations as 
a result of the breakups of large-scale channel flows, similarly to
the no dead zone simulation (SI09). 
The intermittency of the simulated disk winds 
should be observed as the variation of the disk surfaces, which might explain
the observed large time variations of young stars \citep{wis08,muz09,bls09}.

The inside-out clearing of protoplanetary disks by the MRI-driven disk winds 
suppress the infall of boulders because the outward force by gas pressure 
gradient is small. 
This is suitable condition for the formation of planetesimals by aggregation, 
where we also need to examine sticking condition to study 
the actual growth of solid materials \citep[e.g.][]{oku09}. The inside-out
clearing may also increase the dust-to-gas ratio in the inner part of a disk,  
which is also favorable for the formation of planetesimals by gravitational 
instability \citep{sek98,ys02,joh07}.
The migration of planets is also suppressed in 
the inner region where the gas is dispersed at early times. 
Then, the inward migration of newly formed (proto-)planets stop 
at a certain location as shown in Figure \ref{fig:plmg}. 
On the other hand, the gas remains in the outer region, so that the 
formation of gas planets can proceed there.

The authors thank an anonymous referee for many valuable comments. 
This work was supported in part by Grants-in-Aid for 
 Scientific Research from the MEXT of Japan 
 (TKS: 19015004, 20740100, and 22864006 
  SI: 15740118, 16077202, and 18540238), 
 and Inamori Foundation (TKS). 
Numerical computations were in part performed on Cray XT4 at Center for 
Computational Astrophysics, CfCA, of National Astronomical Observatory 
of Japan. 
   



\end{document}